\documentclass[fleqn,usenatbib]{mnras}

\usepackage{newtxtext,newtxmath}

\usepackage[T1]{fontenc}
\defcitealias{Quiros-Rojas2024}{\scshape QR24}

\DeclareRobustCommand{\VAN}[3]{#2}
\let\VANthebibliography\thebibliography
\def\thebibliography{\DeclareRobustCommand{\VAN}[3]{##3}\VANthebibliography}

\usepackage{graphicx}	
\usepackage{amsmath}	

\title[Red-Herschel sources: multiplicity and SFR]{On the multiplicity of red-Herschel sources and its implications for extreme star formation}

\author[
Marianela Quirós-Rojas et al.]{
Marianela Quirós-Rojas$^{1}$\thanks{E-mail: mquiros@inaoe.mx},
Alfredo Montaña$^{1}$,
Jorge A. Zavala$^{2,3}$,
Itziar Aretxaga$^{4,1}$,
    \newauthor
Norma Araceli Nava-Moreno$^{1}$ and David H. Hughes$^{1}$.
\\
$^{1}$Instituto Nacional de Astrof\'{i}sica \'{O}ptica y Electr\'{o}nica, Luis Enrique Erro 1, Tonantzintla CP 72840, Puebla, M\'{e}xico\\
$^{2}$Department of Astronomy, University of Massachusetts, Amherst, MA 01003, USA\\
$^{3}$National Astronomical Observatory of Japan, 2-21-1 Osawa, Mitaka, Tokyo 181-8588, Japan\\
$^{4}$ Centro de Astrobiologia (CAB), CSIC-INTA, Camino Bajo del Castillo s/n, 28692, Villanueva de la Cañada, Madrid, Spain.
\\
}
\date{Accepted XXX. Received YYY; in original form ZZZ}

\pubyear{\the\year{}}

\begin{document}
\label{firstpage}
\pagerange{\pageref{firstpage}--\pageref{lastpage}}
\maketitle

\begin{abstract}
We study the multiplicity of galaxies in the largest sample of red-\textit{Herschel} sources ($S_{250 \mu \mathrm{m}} < S_{350 \mu \mathrm{m}} < S_{500 \mu \mathrm{m}}$) using archival ALMA observations. Out of 2416 fields with ALMA detections (from a total of 3,089 analyzed maps), we identify 474 multiple systems within a radius of 16\,arcsec (equivalent to the 500\,$\mu$m \textit{Herschel} beam-size): 420 doubles, 51 triples, and 3 quadruples. In each case the brightest source contributes, on average, 64, 48, and 42\,per\,cent of the total flux in double, triple, and quadruple systems. The average combined ALMA flux density of the sources in double systems is comparable to that of the two brightest components within triple and quadruple systems. Non-parametric tests suggest that only a small fraction of the double systems ($\lesssim13$\,per\,cent) are comprised of sources with compatible redshifts, while 47-67\,per\,cent of triple and quadruple fields contain at least one potentially associated pair. Simulations using a mock catalogue of dusty star-forming galaxies suggest that 32\,per\,cent of the double systems are likely physically associated ($\Delta z < 0.01$, i.e. $\lesssim$10\,comoving Mpc at $z = 3$) and, while only 8\,per\,cent of the triple and none of the quadruple systems meet this criterion, $\sim$ 70\,per\,cent of them include at least one likely associated pair. Our results suggest that enhanced star formation rates in submillimetre galaxies are primarily driven by internal processes rather than large-scale interactions. This study also provides a catalog of potential overdensities for follow-up observations, offering insights into proto-cluster formation and evolution. 
\end{abstract}




\section{Introduction}
The study of submillimeter galaxies (SMGs), usually identified as the brighter population of dusty star-forming galaxies (DSFGs), began with single-dish telescopes, particularly with the James Clerk Maxwell Telescope (JCMT) and the Submillimetre Common-User Bolometer Array (SCUBA) camera \citep[e.g.,][]{Smail_1997, Hughes1998}, which had an angular resolution of $\sim$15\,arcseconds at 850\,$\mu$m. Subsequent studies revealed key properties of SMGs, including redshift distributions peaking at $z \sim 2-3$ \citep[e.g.,][]{Chapman_2005, Aretxaga_2005, Aretxaga_2007, Yun_2012}, and extreme infrared luminosities ($L_{\mathrm{IR}} \geq 10^{12}$\,L$_\odot$), leading to high star formation rates (SFR $\geq$ 100\,M$_\odot$\,yr$^{-1}$) \citep[see reviews,][]{Blain2002, Casey2014}. However, due to the limited survey areas, it remained challenging to fully characterize the brightest and most extreme members of this population. Large-scale sky surveys are essential to identify and study outliers within the SMG population. The \textit{Herschel Space Observatory} \citep{Pilbratt_2010} is particularly helpful for this purpose, with surveys such as the \textit{Herschel} Astrophysical Terahertz Large Area Survey \citep[\textit{H}-ATLAS;][]{Eales_2010} and the \textit{Herschel} Multi-tiered Extragalactic Survey \citep[HerMES;][]{Oliver_2012}. In particular, \textit{H}-ATLAS covers 660\,square\,degrees of the sky, divided into five fields: GAMA 09, 12, and 15, and the North and South Galactic Poles (NGP and SGP) \citep{Eales_2010}. \textit{Herschel} observed these fields using two continuum cameras: PACS \citep[at 100 and 160\,$\mu$m; ][]{Poglitsch_2010} and SPIRE \citep[at 250, 350, and 500\,$\mu$m;][]{Griffin_2010}. Several studies \citep[e.g.,][]{Hughes2002, Pope_2010, Cox_2011} have shown that SPIRE bands can be used to identify a high-redshift population of SMGs, known as "red-\textit{Herschel} sources", based on the simple criterion $S_{250\mathrm{\mu m}} < S_{350\mathrm{\mu m}} < S_{500\mathrm{\mu m}}$. Given the typical shape of the spectral energy distributions (SEDs) of galaxies’ dust emission, mainly driven by dust temperature ($T_{\mathrm{dust}}$) and emissivity index ($\beta$), this criterion implies that these galaxies lie at redshifts $z>2$.

Red-\textit{Herschel} sources provide a unique laboratory for studying galaxies with extremely high SFR ($\gtrsim$ 1000\,M$_\odot$\,yr$^{-1}$) at high redshifts, offering insights into the most luminous phases of galaxy evolution. However, their nature remains poorly understood, as such extreme properties are challenging to replicate in most theoretical models of galaxy evolution \citep[e.g.,][]{Baugh_2005, Lacey_2010, Hayward_2011,Lacey_2016}. Furthermore, the relatively low angular resolution of \textit{Herschel} \citep[$\theta_{\mathrm{500\mu m}}\sim 37\,$arcseconds;][]{Valiante_2016} limits our ability to fully resolve and characterize this population, as source blending can affect their identification and derived properties \citep[e.g.,][]{Ma_2019,Greenslade_2020,Montaña_2021}.

Advancements in millimeter astronomy, including large single-dish telescopes such as the 30-meter Institute for Radio Astronomy in the Millimeter Range (IRAM)\footnote{\url{https://iram-institute.org/observatories/30-meter-telescope/}} and the 50-meter Large Millimeter Telescope Alfonso Serrano \citep[LMT,][]{LMT}, as well as radio interferometers like the Northern Extended Millimeter Array (NOEMA)\footnote{\url{https://iram-institute.org/observatories/noema/}}, Submillimeter Array \citep[SMA,][]{SMA}, and the Atacama Large Millimeter/submillimeter Array \citep[ALMA,][]{ALMA}, have enabled us to obtain more sensitive observations with relative high angular resolution, reaching sub-arcsecond scales. ALMA, in particular, has demonstrated the ability to resolve multiple sources blended within the typical beam-size of single-dish telescopes. However, most of these follow-up studies have focused on faint SMGs ($S_{\mathrm{1.3\,mm}}<4$\,mJy) identified in single-dish surveys of $\lesssim$1\,square\,degree \citep[e.g.,][]{Hodge_2013, Simpson_2020, dudzevivciute2020,McKinney_2025}.

To overcome the challenges of exploring the most extreme SMGs like red-\textit{Herschel} sources, it is crucial to combine the large-area coverage of \textit{Herschel} with the high angular resolution of ALMA. Characterization of a large sample of these extreme galaxies was presented in \citet[][]{Quiros-Rojas2024} (hereafter referred to as \citetalias{Quiros-Rojas2024}). This study compile the \textit{Herschel}-ALMA Red sources Public Archive Study (HARPAS), which analyzes $\sim$3000 red-\textit{Herschel} sources using public ALMA Band 6 (1.3\,mm) data. It includes 2,416 maps with at least one ALMA detection and 673 maps with non-detections. After excluding 43 detected fields identified as low-redshift galaxies or Active Galactic Nuclei (AGN), the HARPAS catalogue was constructed, comprising 2,373 fields with ALMA detections. Of these, 74\,per\,cent are identified as single sources, 20\,per\,cent are multiple systems, and 6\,per\,cent are gravitationally lensed galaxies and/or close mergers.

The relatively low observed multiplicity (20\,per\,cent) aligns with the findings from other red-\textit{Herschel} sources studies using interferometric observations, where multiplicity estimates range between 20–30\,per\,cent \citep[e.g.,][]{Ma_2019, Greenslade_2020, Bendo, Cox_2024}. These results suggest that multiplicity alone cannot account for the extreme properties of red-\textit{Herschel} sources \citep{Hayward_2013}. Importantly, some of these blended SMGs can serve as tracers of overdensities \cite[e.g.,][]{Hung_2016,Cornish_2024,herwig_2025}, helping to identify progenitors of galaxy clusters as proto-clusters \citep[e.g.,][]{Oteo_2018, Miller_2017}, including overdense regions that are large enough to collapse and eventually form clusters with a total mass of at least $10^{14}$\,M$_\odot$ by $z=0$ \citep{Overzier+2016},  which helps validate cosmological models \citep[e.g.,][]{COORAY20021, Voit_2005, Allen_2011, Kractsov_2012}. However, the nature of these multiple systems remains uncertain without additional spectroscopic information, as some observed systems may be chance projections rather than physically associated structures. 

In this second paper on the HARPAS sample, we present the multiplicity analysis of the largest sample of red-\textit{Herschel} sources with $\sim$1\,arcsecond angular resolution observations. This work aims to enhance our understanding of the role of multiplicity in this extreme submillimeter galaxy population. The paper is structured as follows: Section \ref{sec:HARPAS} provides an overview of the HARPAS catalogue. Section \ref{sec:caractHARPAS} presents the characterization of multiple systems (M-Fields), comparing the findings with simulations of DSFGs. Section \ref{sec:properties} discusses the deblending of SPIRE/\textit{Herschel} fluxes and the physical properties derived through SED fitting. Finally, Section 5 summarizes the results and conclusions.
\par In this work we assume a $\Lambda$CDM cosmology using ${H}_{0}=70\,\mathrm{km}\,{{\rm{s}}}^{-1}\,{\mathrm{Mpc}}^{-1}$, ${{\rm{\Omega }}}_{{\rm{M}}}=0.3$ and  ${{\rm{\Omega }}}_{{\rm{\Lambda }}}=0.7$.

\section{\textit{Herschel}-ALMA Red sources Public Archive Study (HARPAS) catalogue}
\label{sec:HARPAS}
The HARPAS study \citepalias{Quiros-Rojas2024} comprises ALMA observations ($\theta_{\mathrm{FWHM}} \sim 1$\,arcsec, $\sigma_{1.3,\mathrm{mm}} \sim 0.2$\,mJy\,beam$^{-1}$) of more than 3,000 red-\textit{Herschel} sources, identified using the colour criterion $S_{250\,\mu\mathrm{m}} < S_{350\,\mu\mathrm{m}} < S_{500\,\mu\mathrm{m}}$. The sample was constructed by first selecting all red-\textit{Herschel} galaxies in the \textit{H}-ATLAS catalogues \citep[][]{Valiante_2016, Maddox_2018}, resulting in 6,194 sources, which represent 1.4\,per\,cent of all \textit{H}-ATLAS detections. These were then cross-matched with the ALMA public archive using a search radius of 5\,arcseconds, yielding 3,187 red-\textit{Herschel} sources observed across 41 public ALMA projects. Approximately 97\,per\,cent of these (3,089 sources) were observed in Band 6 (1.3\,mm), which provides homogeneous depth and angular resolution; data from other bands were therefore excluded.

\begin{figure}\centering
    \includegraphics[width=0.5\linewidth]{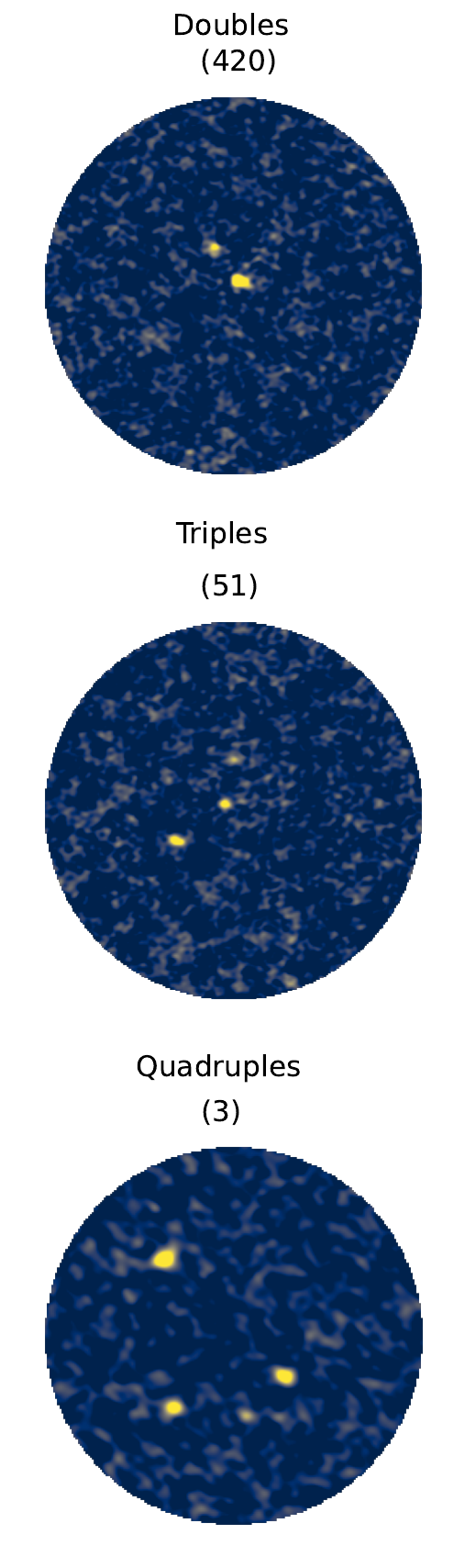}
    \caption{ALMA 1.3\,mm maps showing examples of the three sub-samples in M-Fields (based on the number of sources above 5$\sigma$ within a detection radius of 16.6\,arcseconds): doubles, triples, and quadruples. The values in parentheses indicate the number of fields in each classification out of the 474 M - Fields. The diameter of the ALMA map is $\sim18.6$\,arcseconds}
    \label{fig:types}
\end{figure}

Out of the 3,089 ALMA observed fields, 2,416 include at least one ALMA source detection ($>5\sigma$) and 673 fields with no detections (hereafter N-fields), which are discussed in Section \ref{sec:N-Fields}. Among the ALMA source detected fields, 43 were identified as hosting AGN or low-$z$ galaxies and are excluded from further analysis. The remaining 2,373 fields constitute the HARPAS catalogue, classified in three main groups:

\begin{itemize}
    \item \textbf{S - Fields:} Fields with only a single ALMA detection. This is the dominant category, with a total of 1,762 sources. 
    \item \textbf{PLUM - Fields:} Potential Lenses and Unidentified Mergers (PLUM) are fields with at least two source detections separated by no more than 3\,arcseconds, which may be lensed systems or closely interacting galaxies. The 3 arcsecond criterion was chosen based on the typical separation between images of strongly lensed systems reported in the literature \citep[e.g.][]{Zavala, Gururajan, Bendo}.
    This category contains 137 fields. 
    \item \textbf{M - Fields:} Multiple Fields include those with two or more sources separated by more than 3\,arcseconds. Fields containing at least one source separated by more than 3\,arcseconds, regardless of the presence of other sources closer than 3 arcseconds, are included in this classification. This category contains 474 systems, of which 420 have only two components, 51 have three, and 3 fields have four (see examples in Fig. \ref{fig:types}). 
\end{itemize}

\begin{figure}
    \centering
    \includegraphics[width=\linewidth]{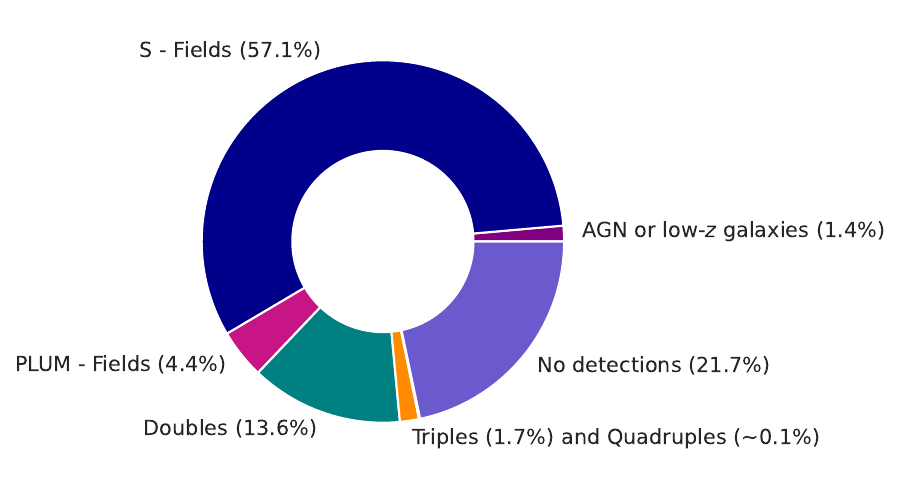}
    \caption{Percentages of the different subsamples of the $\sim$3,000 red-\textit{Herschel} sources classified from the 1.3\,mm ALMA maps. The HARPAS catalogue contains three main categories: S-Fields, PLUM-Fields, and M-Fields. The M-Field classification is further subdivided into doubles, triples, and quadruples.  }
    \label{fig:all_sample}
\end{figure}

 The classification of the entire sample is shown in Figure \ref{fig:all_sample}. The first two sub-catalogues, which contain the individual sources (S-Fields) and Potential Lenses and Unidentified Mergers (PLUM-Fields) of the HARPAS sample were presented and analyzed in \citetalias{Quiros-Rojas2024}. Here we expand analysis of the HARPAS catalogue by characterizing sources from the M-Fields (see details below in Section \ref{sec:M-fields}). The catalogs provide deblended flux density measurements from the SPIRE/\textit{Herschel} bands at 250, 350, and 500\,$\mu$m, as well as ALMA 1.3\,mm photometry.

\section{Multiplicity of red-\textit{Herschel} sources}\label{sec:caractHARPAS}
Studying the M-Fields provides valuable insights into the nature of red-\textit{Herschel} galaxies. ALMA maps allow us to study multiplicity within a \textit{Herschel} beam, as their sizes are comparable to the \textit{Herschel} beam at 500\,$\mu$m ($\sim 37$\,arcseconds). Furthermore, due to ALMA's  angular resolution, we are sensitive to compact sources separated by more than $\sim 1$\,arcsecond.

In this section, we consider the nature of multiplicity in observations of red-\textit{Herschel} sources by characterizing flux densities and angular separations between multiple-systems components. Furthermore, we investigate how instruments at different millimeter-wavelengths observatories are capable of resolving these multiplicities. 

\subsection{Multiplicity fraction}
\label{sec:M-fields}

From our sample of 2416 red-\textit{Herschel} sources with 1.3 mm ALMA detections, we find a multiplicity fraction of 20\,per\,cent (i.e., 474 fields). This value is slightly lower than those reported in some previous studies, but remains consistent within the uncertainties. However, as discussed below, those earlier estimates were based on significantly smaller samples, which may be more affected by uncertainties and biases.
In addition, the 3\,arcseconds separation limit adopted to distinguish between PLUM-fields and M-fields (Sec. \ref{sec:HARPAS}) implies that a small fraction of PLUM-Fields may actually be short-separation multiple systems (i.e. M-Fields). If we assume that all PLUM-Fields are multiple systems, the multiplicity fraction would increase from 20 to 25\,per\,cent. What follows is a brief comparison between our results and previous results from the literature.

\citet{Greenslade_2020} studied a sample of 34 red-\textit{Herschel} sources from the HERMES survey with SMA observations ($\theta_{\mathrm{FWHM}} \sim 0.35-3$\,arcseconds),of which 24 have detections, yielding a multiplicity of $\sim$12\,per\,cent (i.e., four fields with multiple detections), three of which have $S_{500\mathrm{\mu m}} > 60$\,mJy. Furthermore, \citet{Cairns_2022}, using deeper SMA observations, showed that three of the four galaxies undetected by \citet{Greenslade_2020} are resolved into multiple components. Moreover, \citet{Montaña_2021}, studying multiplicity of a sample $\sim$100 DSFG at scales smaller than the 32\,m-LMT AzTEC beam ($\sim9.5$\,arcseconds), conducted two tests to compare the measured PSF profile of single sources with that of an expected point source. Deviations in the width and shape of the PSF suggested the presence of two or more sources blended within the AzTEC beam. Their analysis revealed a multiplicity rate of approximately 20\,per\,cent. Finally, \citet{Ma_2019} using ALMA, NOEMA, and SMA observations of 300 red-\textit{Herschel} sources---the largest sample followed-up with interferometric observations before the HARPAS survey, found that 27\,per\,cent (i.e., 81 sources) were resolved into multiple components.

The results from the HARPAS sample, along with those reported in the literature, confirm the relatively low fraction of multiplicity present in red-\textit{Herschel} sources, which supports the idea that multiplicity is not the primary explanation for their high brightness \citep[e.g.,][]{Hayward_2013,Quiros-Rojas2024} and hence they are intrinsically very luminous sources in the early universe. This remains true even in the case where all the non-detections are assumed to be multiple systems as discussed below in Sec. \ref{sec:N-Fields}.

On the other hand, while the estimation of the source number counts is beyond the scope of this work, our analysis indicates that the combined effect of multiplicity and flux boosting introduces a maximum factor of $\sim$2.6 overestimation in the number of $S_{500\mathrm{\mu m}} > 40$\,mJy galaxies in our sample. This effect decreases with increasing $S_{500\mathrm{\mu m}}$ limit, with a $\sim$2.1 overestimation for sources with $S_{500\mathrm{\mu m}} > 80$\,mJy and with consistent results at $S_{500\mathrm{\mu m}} > 120$\,mJy. While this demonstrates the importance of considering the effect of multiplicity when estimating and interpreting the number counts from \textit{Herschel} observations, we emphasize that our results are derived from a colour-selected sample, and may not accurately represent the larger population of \textit{Herschel} sources, since the lower redshift population (i.e. non red-\textit{Herschel} sources) might have a different multiplicity fraction.

\subsection{Flux Density Scaling with Multiplicity}
\begin{figure}
    \centering
    \includegraphics[width=\linewidth]{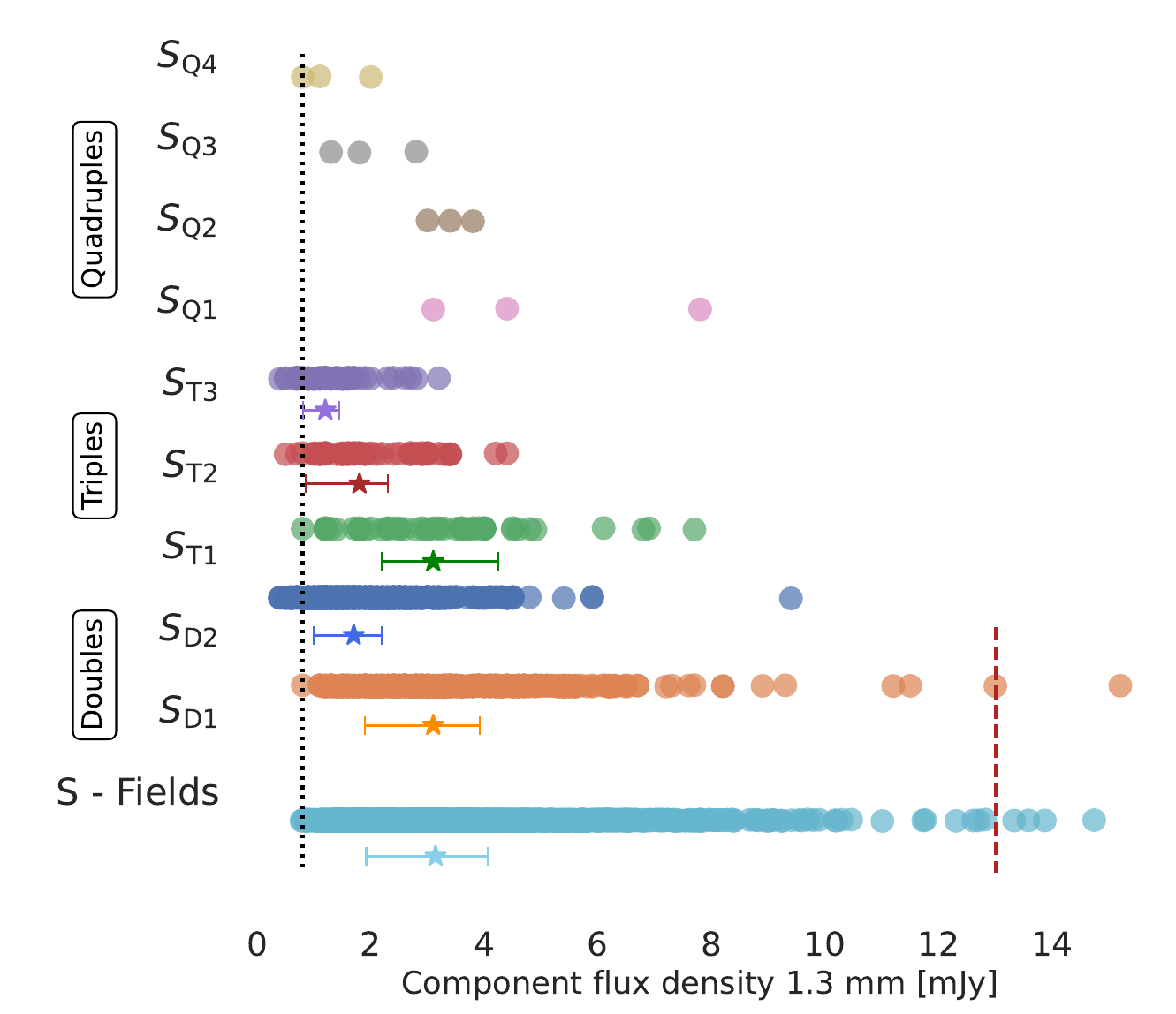}
    \caption{Flux density of each component in multiple systems compared to the S-Fields \citepalias{Quiros-Rojas2024}. Components are denoted by D$_i$ for doubles, T$_i$  for triples, and Q$_i$  for quadruples. The subindex, $i$, indicates the relative brightness of each component, with 1 for the brightest source. Stars denote the median flux densities, while the error bars represent the 1st and 3rd quartiles of the flux density distribution. The dotted black line indicates the 5$\sigma$ median 1.3\,mm detection threshold for the full HARPAS sample, while the dashed red line represents the flux limit proposed by \citetalias{Quiros-Rojas2024}, above which 100\,per\,cent of the SMGs are gravitationally amplified. As can  be seen, the brighter components ($S_{\mathrm{D1,T1,Q1}}$) in each system (e.g., doubles, triples, and quadruples) exhibit a median flux density comparable to the S-Fields.}
    \label{fig:all_fluxes}
\end{figure}

\begin{figure}
    \centering
    \includegraphics[width=\linewidth]{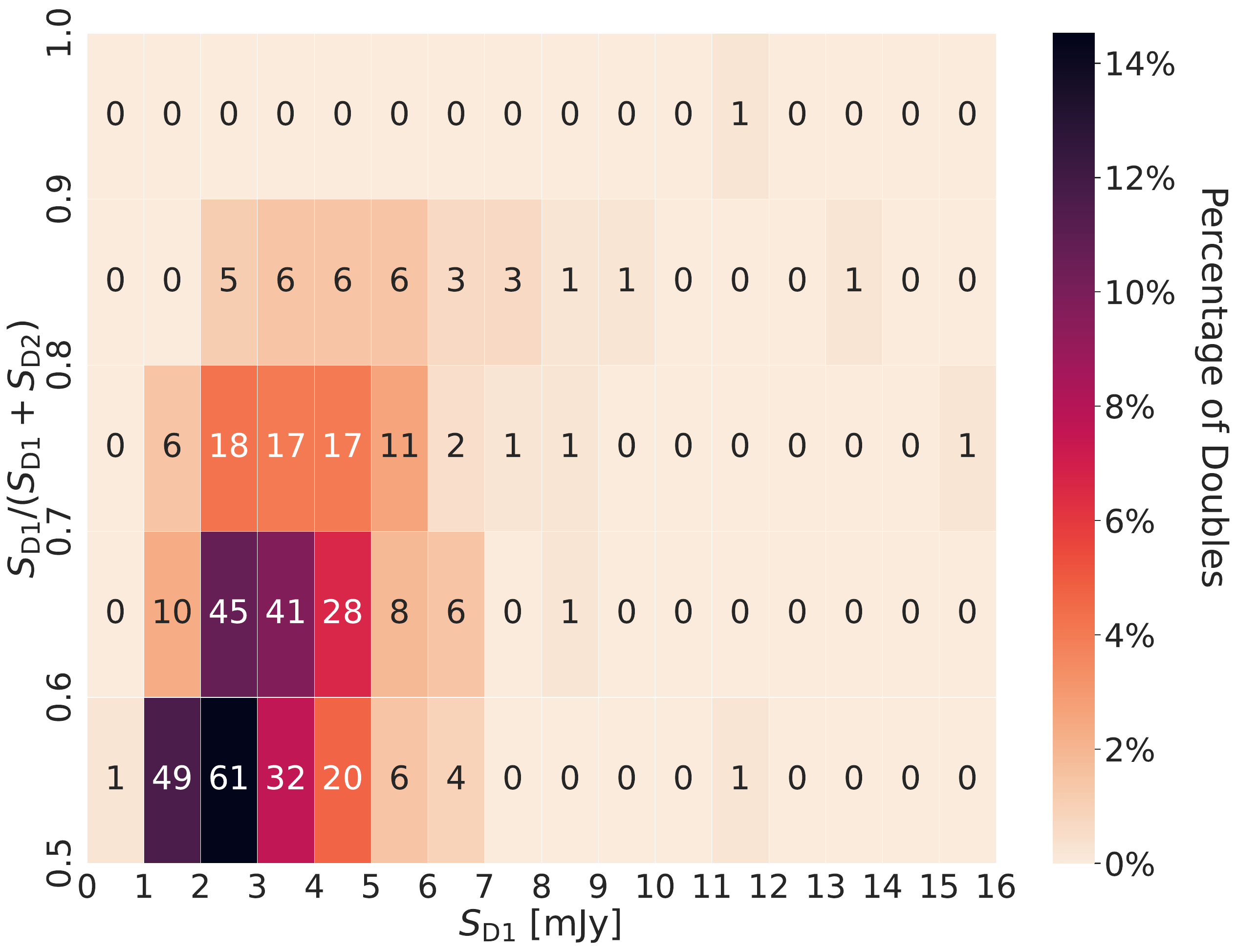}
    \caption{Flux density of the brighter component in each double system ($x$-axis) and its contribution to the total flux density of the system ($y$-axis). The numbers inside the rectangles represent the number of doubles within each flux density bin. The color bar indicates the percentage of doubles in each bin. On average, the brighter component contributes approximately 64\,per\,cent of the total flux across most doubles in each field.}
    \label{fig:frac_pairs_1}
\end{figure}

\begin{figure*}
    \centering
    \includegraphics[width=\linewidth]{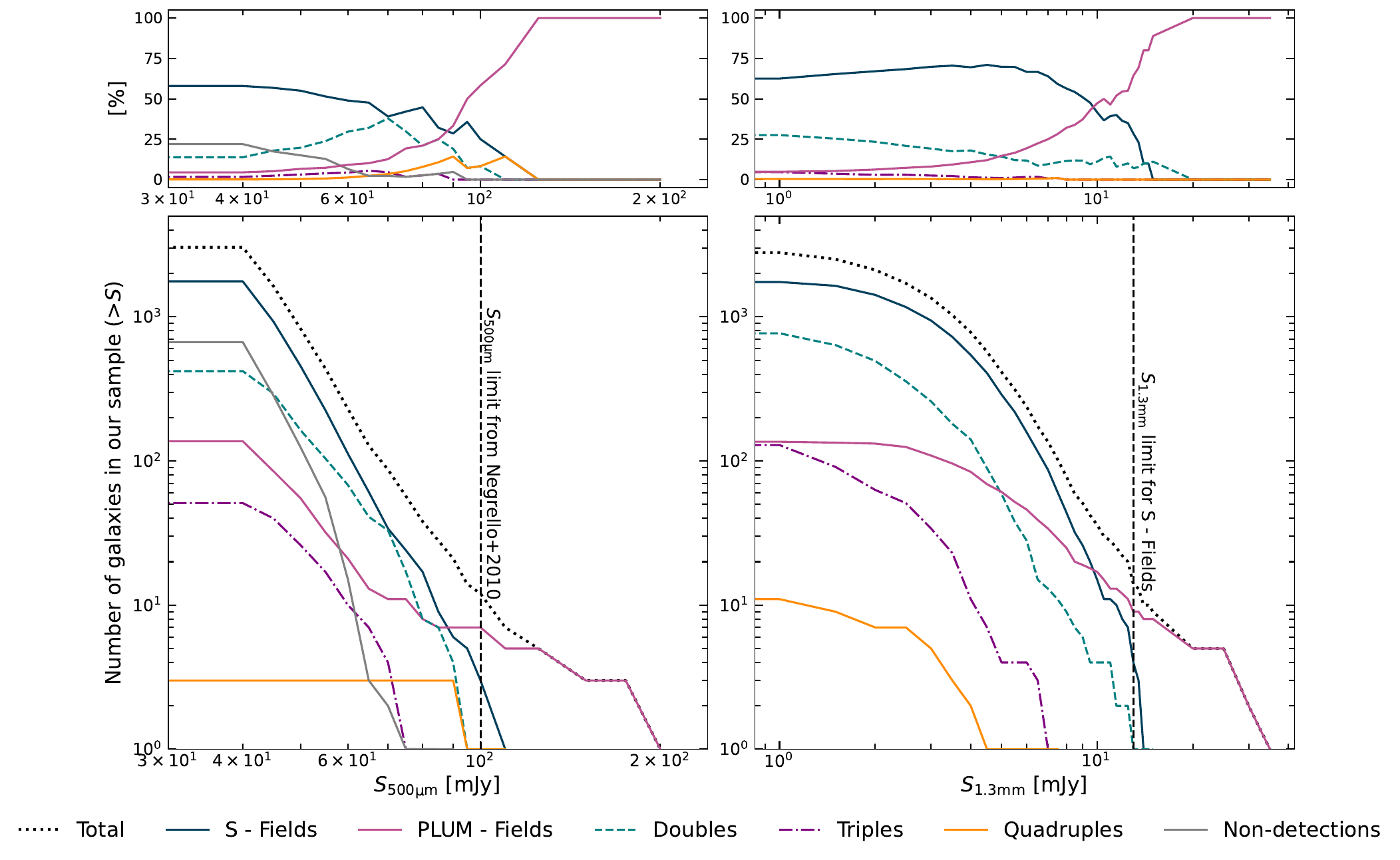}
    \caption{Cumulative number of galaxies (\textit{bottom panels}) for the HARPAS sample as a function of \textit{Herschel}/SPIRE 500\,$\mu$m (\textit{left}) and 1.3\,mm (\textit{right}) flux density and split according to their classification. For the 1.3\,mm panel, we include all the individual components within each system, while in the 500\,$\mu$m panel we use the original flux density of each field reported in the \textit{H}-ATLAS catalogue. The top panels, show the fractional contribution to the total of red-\textit{Herschel} fields observed with ALMA (\textit{left}) and to the total of ALMA detections (\textit{right}). Fields with no detections in ALMA tend to be fainter at 500\,$\mu$m. We also indicate the threshold proposed in \citetalias{Quiros-Rojas2024} ($S_{1.3\mathrm{mm}} \geq 13.0$\,mJy), for identifying  gravitationally lensed candidates. This limit remains applicable to multiple systems, as most galaxies in these systems have flux densities below this threshold, with the exception of one, which may exhibit signs of gravitational amplification below our angular resolution ($\sim1$\,arcsecond).  }
    \label{fig:number_sources}
\end{figure*}

\begin{figure*}
    \centering
    \includegraphics[width=\linewidth]{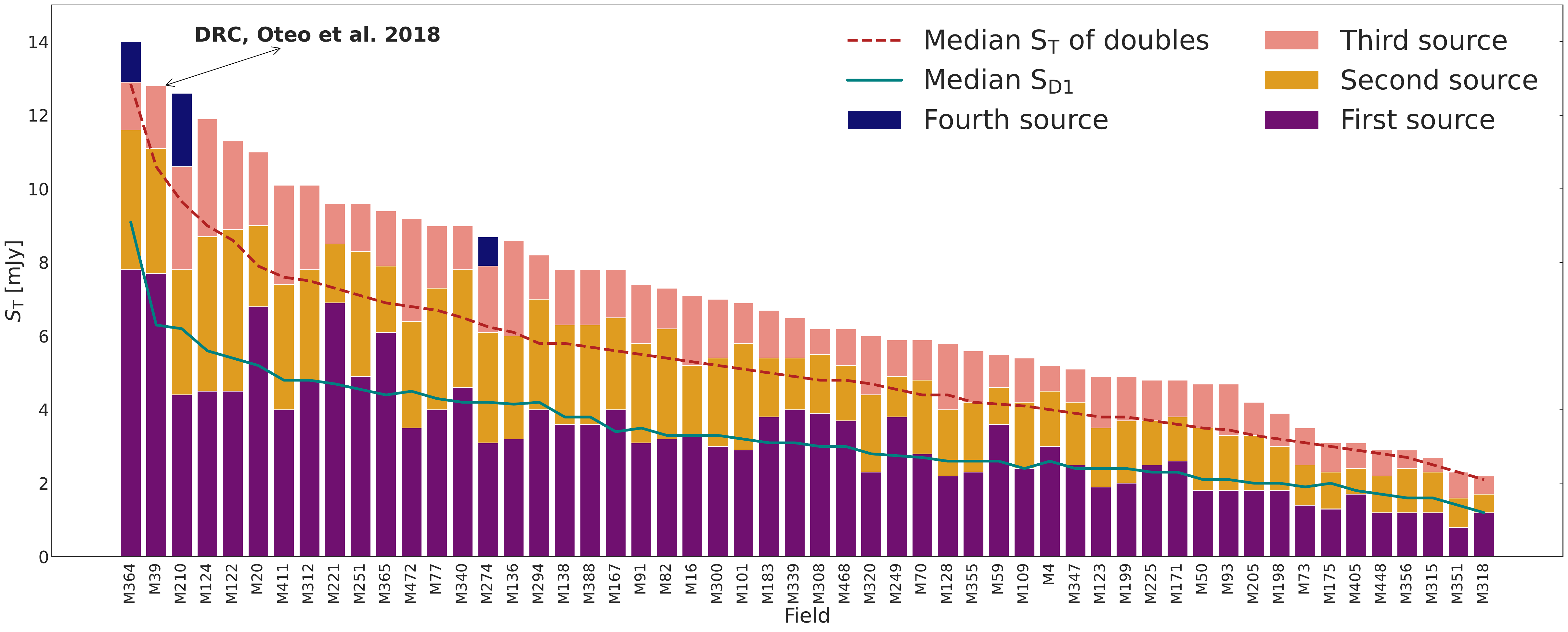}
    \caption{The 51 triple systems and the 3 quadruple systems shown in order of decreasing total flux density (S$_{\mathrm{T}}$). The contribution to S$_{\mathrm{T}}$ from the different components are shown with different colors. For comparison, we also show  the average total flux density (dashed red line) and the flux density of the brightest component (solid green line) of the double systems, computed by generating 100 random subsamples (see details in the main text). The total flux density of the doubles are, on average, fainter than most of the triple or quadruple fields, but consistent with the total flux density of their two brighter components. 
    Additionally, we highlight the proto-cluster DRC \citep{Oteo_2018}, which is part of our sample and classified as a triple. This overdensity has been confirmed to be a proto-cluster, containing at least ten DSFGs at $z=4.002$ over a $\sim 0.44$ square arcminutes area.}
    \label{fig:frac_t_q_1}
\end{figure*}

Figure \ref{fig:all_fluxes} presents the flux density for each component in the S-Fields, doubles, triples, and quadruples. We calculate the median flux density ratios between the brightest and the fainter components of the systems: 1.8 for doubles, 1.7 and 2.6 for triples, and 1.3, 2.4, and 4.0 for quadruples. Interestingly, the median flux density of the brightest components in the doubles and triples is comparable to the median flux density observed in the S-Fields, as clearly seen in figure \ref{fig:all_fluxes}.

We find that the median contribution of the brightest source to the total flux in the double, triple, and quadruple systems is 64, 48, and 42\,per\,cent, respectively (Fig. \ref{fig:all_fluxes}). We further illustrate this in Figure \ref{fig:frac_pairs_1}, where we explore the relative contribution of the brightest source in the double systems as a function of flux density. The brightest component of approximately 70\,per\,cent of these double systems have 1.3\,mm flux densities between 1 and 5\,mJy, corresponding to a 50-70\,per\,cent contribution to the total flux. These results are consistent with those reported by \citet{Ma_2019}, who studied a sample of 300 \textit{Herschel} ultra-red DSFGs using data from IRAC/\textit{Spitzer} at 3.6 and 4.5\,$\mu$m, as well as observations from ALMA, NOEMA, and SMA. Their study found that the brightest components of these ultra-red \textit{Herschel} sources, observed with ALMA, contribute 41–80\,per\,cent of the total ALMA flux at 870\,$\mu$m and 3\,mm.  For the double systems, in particular, the brighter source contributes between 50 to 80\,per\,cent.

One of the brightest galaxies in the entire HARPAS catalogue is found within the doubles sample (see Fig. \ref{fig:all_fluxes}). This source, HARPAS\_1486 with a flux density of $15.2\pm2.1$\,mJy, exceeds the 1.3\,mm flux density threshold proposed by \citetalias{Quiros-Rojas2024}, above which galaxies are likely to be gravitationally lensed.  This raises the possibility that the brightest source in our M-Field sample is also subject to gravitational lensing. Furthermore, the deblended flux density at 500\,$\mu$m is above 90\,mJy, which is close to the limit proposed by \citet{negrello_2010} for confidently identifying lensed sources. However, the ALMA map with an angular resolution of approximately 1.7$\times$1.3\,arcseconds, does not resolve any potential lensing features. Its measured flux density lies close to the classification threshold when accounting for measurement uncertainties, preventing a confident identification as a lensed source. Further observations with higher angular resolution are required to confirm the nature of this system.

Figure \ref{fig:number_sources} presents the cumulative number of galaxies in the HARPAS sample as a function of the 500\,$\mu$m flux density (taken from the \textit{H}-ATLAS catalogue)  and as a function of the ALMA 1.3\,mm flux density, along with their fractional contribution to the total sources. With the exception of the source discussed above (HARPAS\_1486), sources in the M-fields do not exceed the 1.3\,mm flux density nor the 500\,$\mu$m thresholds proposed by \citetalias{Quiros-Rojas2024} and \citet{negrello_2010}, above which galaxies are considered to be affected by gravitational lensing.

Figure \ref{fig:frac_t_q_1} shows the total flux density of all 54 fields containing triple and quadruple systems. To compare these with double systems, we use the full set of doubles (420 fields), from which we randomly select 54 fields to match the sample size of the triples and quadruples. This procedure is repeated 100 times. For each subsample, we compute the median total flux density and the flux of the brightest component, after sorting the sources by brightness. On average, the total flux density of the double systems is lower than that of the triple and quadruple systems, but it is consistent with the combined flux of their two brightest components.

\subsection{Angular Separation of Multiple Systems}
\begin{figure}
    \centering
    \includegraphics[width=\linewidth]{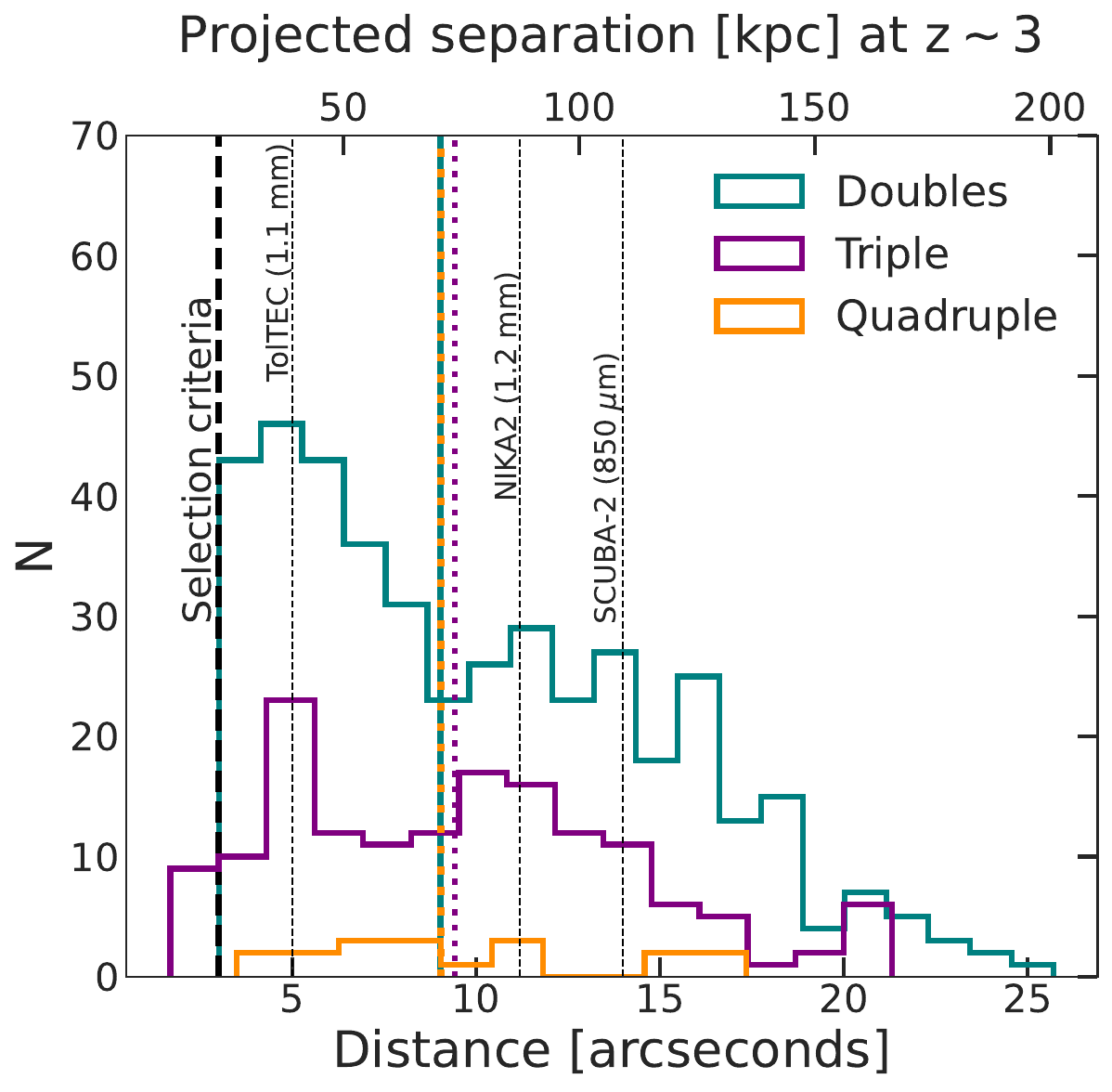}
    \caption{Apparent separation between each pair of sources for the doubles, triples, and quadruples, with medians of 9.03, 9.42, and 9.04\,arcseconds, respectively. The triples show separations below our selection criteria for the M-Fields (dashed black line), because if a source has a distance larger than three arcseconds, it will be classified as an M-Field even if a closer source exists. Above 16.6\,arcseconds (the detection radius), our sample is affected by the decreasing sensitivity of ALMA with increasing map radius. Additionally, the black dashed lines represent the angular resolution of TolTEC \citep{toltec}, NIKA2 \citep{nika2}, and SCUBA-2 \citep{scuba-2}. }
    \label{fig:distance}
\end{figure}

The distances between components may offer insights into the physical processes that can enhance the SFR. Figure \ref{fig:distance} shows the distances between components within each sub-sample. Since our detections were made within a radius of 16.6\,arcsecond \citepalias[][]{Quiros-Rojas2024}, the maximum possible separation between the components is twice this value.  All three sub-samples exhibit a median distance of approximately 9 arcseconds between components, with doubles reaching up to 25 arcseconds. 

To investigate whether doubles with approximately equal flux density components could result from interactions, we examined their angular separations and found that such sources are distributed along the full range of distances (3-25\,arcseconds, Fig. \ref{fig:distance}), indicating that there is no correlation between separation and flux density. Although the angular separation could provide some insight into whether the SFR is enhanced by interactions, it is important to determine whether these sources are physically associated. This will be discussed in the following sections, but we note that doubles and triples display a similar bimodal distribution of distances, with peaks around 5 and 12 arcseconds (although this could also be driven by statistical noise  or by the binning of the data).

Characterizing the typical separation of the multiple systems allows us to investigate the possibility of observing this multiplicity with single-dish telescopes. To do so, we compare the distances between galaxies with the angular resolutions of TolTEC on the 50\,m LMT \citep[$\theta_{\mathrm{FWHM}} \sim 5\,\mathrm{arcseconds}$ at 1.1\,mm;][]{toltec}, NIKA2 on the 30\,m IRAM telescope \citep[$\theta_{\mathrm{FWHM}} \sim 11.2\,\mathrm{arcseconds}$ at 1.2\,mm;][]{nika2}, and SCUBA-2 on the 15\,m JCMT \citep[$\theta_{\mathrm{FWHM}} \sim 14\,\mathrm{arcseconds}$ at 850\,$\mu$m;][]{scuba-2}. These instruments would resolve approximately 82, 40, and 24\,per\,cent of the doubles, and 47, 4, and 2\,per\,cent of the triples as three-component systems. For quadruple systems, TolTEC and NIKA2 could resolve only one of these systems, while SCUBA-2 would not be able to resolve any of them. However, these results do not take into account the confusion limit of each instrument, which may prevent the detection of these galaxies.

\section{Physical Properties}\label{sec:properties}
\subsection{Deblending of flux density in \textit{Herschel} maps}\label{sec:deblending}
Due to the relatively low angular resolution of the \textit{Herschel} maps ($\sim$18, 25, and 36 arcseconds at 250, 350, and 500 $\mu$m, respectively), sources separated by only a few arcseconds would be blended within the \textit{Herschel} beams.

To address this limitation, we use \textsc{XID+} \citep{XID+}, which provides full posterior flux distributions for each deblended component, yielding more reliable flux density estimates and a clearer evaluation of blending effects.

 \texttt{XID+} is a software that employs Bayesian probabilistic modeling. This approach utilizes prior information from the observed data and provides a full posterior distribution of the flux densities. For this analysis, we use the positions derived from the 1.3\,mm ALMA data as prior information and adopt the default uniform prior range of 0.01 to 1000\,mJy.

To estimate uncertainties of the deblended flux densities, we use the methodology presented by \cite{Wang_2024}. Specifically, we calculate the uncertainty for each SPIRE band by determining the 16th and 84th percentiles of the flux density distribution, estimate the difference between these values and the median flux density, and take the larger of these differences as the flux density uncertainty. This value is then combined in quadrature with the residual noise component from confusion (extracted using \texttt{XID+}) and the calibration error, assumed to be 5.5\,per\,cent \citep{Valiante_2016}. The median uncertainties at 250 $\mu$m, 350 $\mu$m, and 500 $\mu$m are 8.1, 9.4, and 11.4 mJy for the doubles; 8.4, 10.6, and 11.5 mJy for the triples; and 6.9, 13.3, and 17.4 mJy for the quadruples.

\begin{figure}
    \centering
    \includegraphics[width=\linewidth]{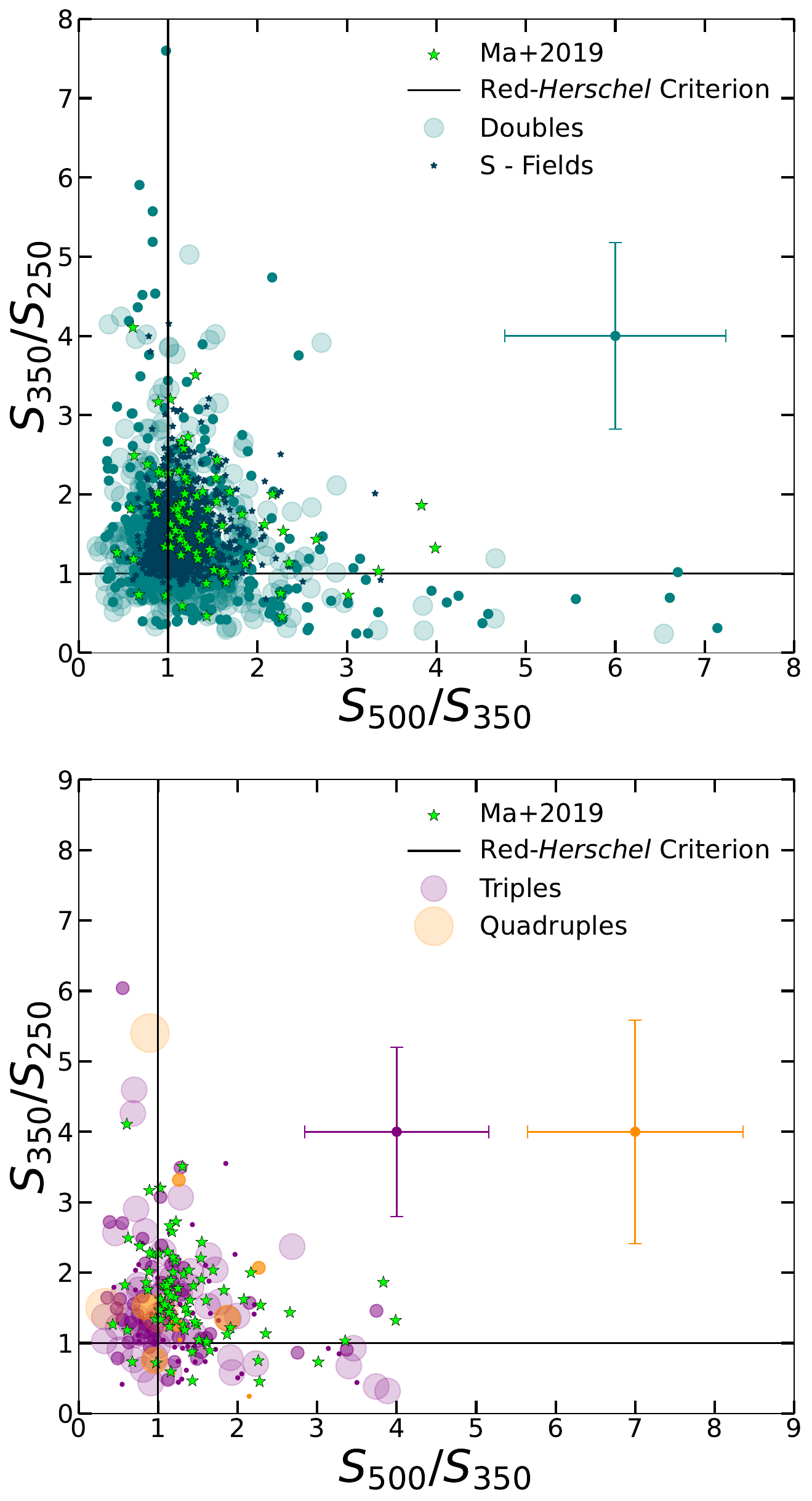}
    \caption{Colour-colour plot of the SPIRE/\textit{Herschel} bands (after deblending) of the red-\textit{Herschel} sources in our M-field sub-sample: doubles (teal circles), triples (purple circles) and quadruples (orange circles). The size of the circles represents the relative brightness of each galaxy compared to other sources in the same field, with the largest circle indicating the brightest source, while the lines represent the red-\textit{Herschel} source criteria ($S_{250 \mu \mathrm{m}} < S_{350 \mu \mathrm{m}} < S_{500 \mu \mathrm{m}}$). The error bars indicate the median uncertainties in the colours for the doubles (top panel), and for the triples and quadruples (bottom panel).} For comparison, the ultra-red sample from \citet{Ma_2019} is shown as green stars. After applying deblending techniques as \texttt{XID+}, 30\,per\,cent of the fields contain sources that no longer meet the red-\textit{Herschel} criterion.
    \label{fig:colour}
\end{figure}

Figure \ref{fig:colour} shows the distribution of M-Fields in the SPIRE colour–colour diagram following deblending. After this step, 70\,per\,cent of the fields contain at least one source that satisfies the red-\textit{Herschel} criterion. Similarly, using \texttt{XID+} to analyze a sample of ultrared DSFGs ($S_{500 \mu \mathrm{m}} / S_{250 \mu \mathrm{m}} \geq 1.5$ and $S_{500 \mu \mathrm{m}} / S_{350 \mu \mathrm{m}} \geq 0.85$) from H-ATLAS and HerMES, \citet{Ma_2019} found that 70\,per\,cent of their sources remained classified as red-\textit{Herschel} after deblending. This comparison highlights that while most of these galaxies retain the red-\textit{Herschel} classification, a non-negligible fraction does no longer meet this criterion.

\subsection{Redshift, IR Luminosities and Star Formation Rates}
To estimate photometric redshift ($z_{\mathrm{phot}}$) and infrared luminosities ($L_{\rm IR}$) from the available photometry, we use \texttt{MMPz} \citep{Casey_2020}. This software fits a far-IR millimeter modified blackbody spectrum with $\beta=1.8$ and a mid-IR power-law model ($\alpha_{MIR}=3$) to the data points (in this case the deblended flux density from \textit{Herschel}/SPIRE and the ALMA 1.3\,mm photometry; see Section \ref{sec:HARPAS}). \texttt{MMPz} estimates the best-fit parameters by performing SED fitting, which derives the $z_{\mathrm{phot}}$ and $\lambda_{\mathrm{peak}}$, the peak wavelength of thermal dust emission. $\lambda_{\mathrm{peak}}$ is related to the dust temperature, which in turn is linked with the $L_{\rm IR}$. This process generates a two-dimensional distribution with a redshift track, which traces how the SED fitting changes as a function of the redshift. This is then compared with the redshift-independent relation between $\lambda_{\mathrm{peak}}$ and $L_{\rm IR}$, as outlined in \citet{Casey_2018}.

\begin{figure*}
    \centering
    \includegraphics[width=\textwidth]{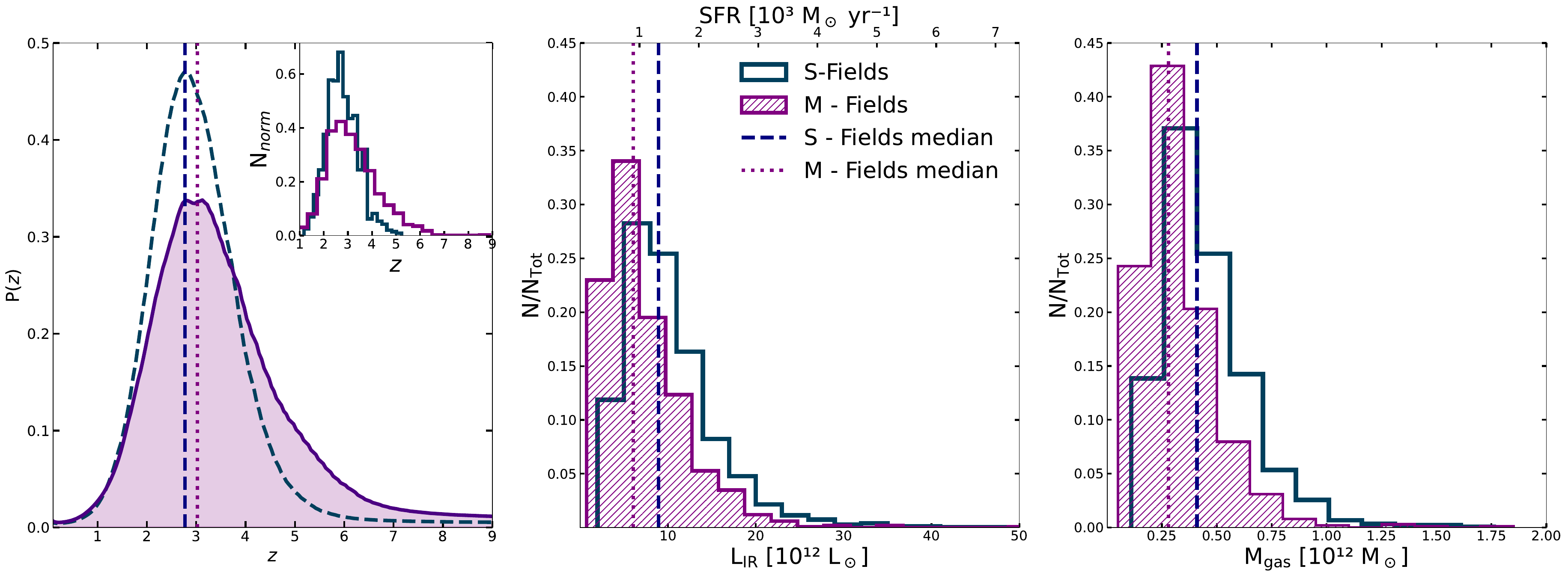}
    \caption{Histograms of the physical properties estimated for all the ALMA sources in the M-Fields using 1.3\,mm ALMA and deblended \textit{Herschel}/SPIRE photometry: redshift probability distribution (left) and histograms of the best-fit photometric redshifts (inset), infrared luminosities and SFR (middle), and gas mass (right). The median values of the distributions are indicated with purple dotted lines and, for comparison, those for S-Fields presented in \citetalias{Quiros-Rojas2024} are indicated with blue dashed lines. This shows that sources in multiple-component systems are less luminous compared to the S-Fields; however, their redshift probability distributions are similar, with the former exhibiting a more pronounced tail at $z > 4$.}
    \label{fig:properties}
\end{figure*}

Figure \ref{fig:properties} shows the properties estimated using \texttt{MMPz}, along with the SFR, which is calculated using the relation presented in \citet{Kennicutt_2012}: $
\mathrm{SFR} \, [\mathrm{M_{\odot}} \, \mathrm{yr}^{-1}] = 1.48 \times 10^{-10} \, L_{\mathrm{IR}} \, [\mathrm{L_{\odot}}],$ which assumes a \citet{Kroupa_2001} Initial Mass Function. Molecular gas masses are derived from the continuum data at 1.3\,mm using the method described by \citet{Scoville_2016}. 

For all the ALMA sources in M-Fields, we obtain a median $z_{\mathrm{phot}}$ of 3.0 [2.2–4.2], where the values in brackets represent the range that encompasses 68\,per\,cent of the distribution (i.e., the 16th/84th percentiles). The median lower and upper uncertainty limits in the photometric redshifts are 0.35 and 0.75, respectively. This agrees with the population of S-Fields in \citetalias{Quiros-Rojas2024}, which has a median value of 2.78 $\pm$ 0.03, similar to those of the general DSFG population with median values typically ranging between $z\sim$2.5 and 3 \citep[e.g.,][]{HATSUKADE_2018, Franco_2018, Simpson_2020, Dudze_2020}. These values are lower than those reported in \citet{Montaña_2021}, who studied a sample of $\sim$100 ultra-red \textit{Herschel} sources with 9.5 arcsec resolution AzTEC/LMT observations and finding that those sources which split into multiple components exhibited a lower mean redshift ($z\sim3.5$) than single systems ($z\sim3.8$). Nevertheless, their more extreme colour selection might contribute to the inferred higher redshifts.

\begin{figure}
    \centering
    \includegraphics[width=\linewidth]{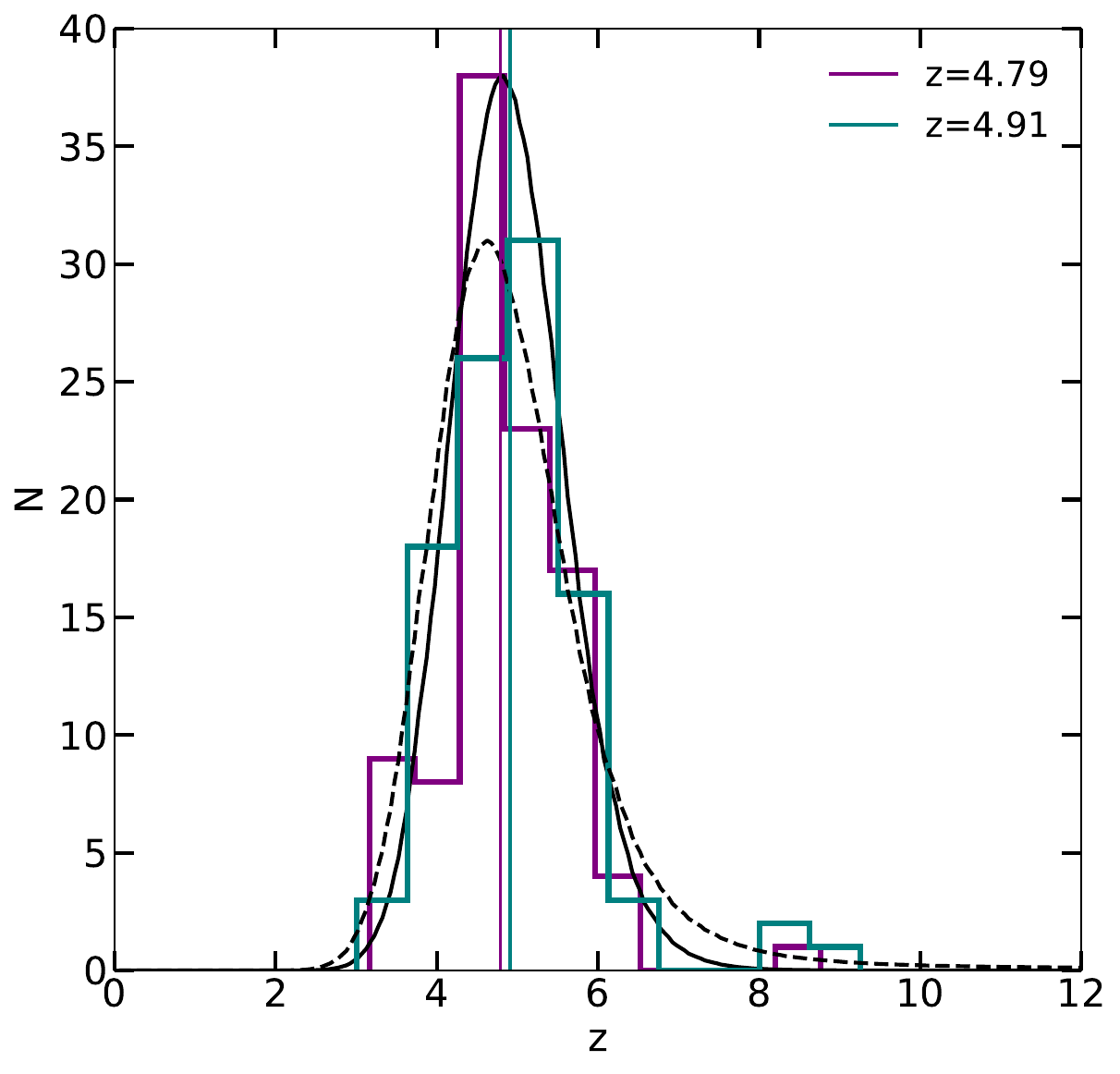}
    \caption{Histograms in teal and purple show the redshift samples generated to follow the redshift probability distributions for the two sources in the M267 field (i.e., HARPAS\_1748 and HARPAS\_1749), classified as a double. The solid and dashed grey lines represent the redshift probability distributions obtained from \texttt{MMPz}. A Mann–Whitney U test yields a $p$-value greater than 0.05 ($p=0.23$), indicating that the median redshifts are statistically consistent with being drawn from the same distribution. For the 13\,per\,cent of pairs that the Mann–Whitney U test suggests are physically associated, the median $\Delta z$ is 0.5.}
    \label{fig:MW}
\end{figure}

Given the substantial uncertainties associated with IR–mm photometric redshifts (typically $\Delta z\approx1$ for our galaxies), which are larger than those derived from optical/near-infrared observations and further increased by the deblending process in M-Fields, we are unable to draw strong conclusions regarding the physical association of the multiple systems in our sample. However, we tentatively explored this possibility by comparing their median redshifts using non-parametric statistical tests, specifically the Mann–Whitney U and Kruskal–Wallis tests. We generated 100 redshifts per source that follow MMPz redshift probability distribution (see Fig. \ref{fig:MW}), and we applied the Mann–Whitney U test to assess whether the two sources have median redshifts that are compatible with a common distribution. Hence, we propose candidate associations with this analysis. For triple and quadruple fields, we conducted: 1) pairwise Mann–Whitney U tests for all source combinations; and 2) the Kruskal–Wallis test, which generalizes the comparison to more than two medians and their distributions. High $p$-values (e.g. $p > 0.05$) indicate that the median redshift of the sources are statistically compatible with arising from a common parent distribution.

For the doubles, 13\,per\,cent of the sources have median redshifts that are compatible with a common redshift distribution. For the triple and quadruple fields, the pairwise Mann–Whitney U tests indicate that 47 and 67\,per\,cent of the fields, respectively, contain at least one pair of sources with median redshifts compatible with a common redshift distribution. Moreover, the Kruskal–Wallis test yields $p$-values above 0.05 in only 2\,per\,cent of the triple fields and in none of the quadruple fields. These results provide tentative insight into the likelihood of physical association among sources within each field while emphasizing that only photometric redshifts are available, with significant associated uncertainties. It is important to note that although simulations and statistical tests suggest a low probability of physical association, some of these multiplicities may still have a common origin \citep[e.g. DRC;][]{Oteo_2018}. However, definitive conclusions cannot be drawn until spectroscopic redshifts are available.

The median $L_{\mathrm{IR}}$ of the ALMA detections is $6.0 [3.1 - 11.1] \times 10^{12}$ L$_{\odot}$, with a 68th percentile indicated in brackets, and a median SFR of $\sim 894 \, [470-1640]$\,M$_{\odot}\, \mathrm{yr}^{-1}$. This value is slightly lower than the 1300$\pm30$ M$_{\odot}\, \mathrm{yr}^{-1}$ found for S-Fields \citepalias{Quiros-Rojas2024}. These intrinsically fainter sources still exhibit properties characteristic of starburst galaxies, such as extreme star formation \citep[e.g.,][]{Ma_2019,Montaña_2021}. The maximum SFR for the M-Fields is $\sim$5,000\,M${_\odot}\,\mathrm{yr}^{-1}$, which is consistent with the SFR limit found in \citetalias{Quiros-Rojas2024} for S-Fields ($\sim$6,000\,M${_\odot}\,\mathrm{yr}^{-1}$). 
The median total SFR, defined as the sum of the SFRs for all ALMA sources within each field, is $\sim$2000, 2900, and 3700\,M${_\odot}\,\mathrm{yr}^{-1}$ for doubles, triples, and quadruples, respectively.

Finally, we estimate a median molecular gas mass of $2.8 \times 10^{11} \, \mathrm{M_{\odot}}$, using \citet{Scoville_2016} and the 1.3 mm data,  with the 68th percentile range of $[1.7-4.7] \times 10^{11} \, \mathrm{M_{\odot}}$. These values are consistent with molecular gas masses estimated using different CO transitions for samples of a few to tens of infrared luminous DSFGs \citep[e.g.][]{bothwell,Aravena_2016,yang_2017,Birkin_2020,Ikarashi_2022,Berta_2023}. These systems exhibit molecular gas masses that closely resembling those found in the S-Fields. These results suggest that, even if these galaxies are associations, they are not fundamentally different from more isolated galaxies.

\section{Comparison with simulations} \label{sec:sim}

We compare our observations against an extended version of the mock redshift survey produced by \citet{Nava_Moreno_2024}, which models DSFGs over a 100 square degree area, including the brighter population of SMGs. This simulation is based on the MultiDark-\textit{Planck} 2 dark matter halo simulation, with infrared properties assigned through a combination of theoretical and empirical relations. 
In particular, the adopted dust-obscuration fractions ($f_{\mathrm{obs}}$) are based on the empirical relation between $f_{\mathrm{obs}}$ and stellar mass from \citet{Whitaker_2017}, and extrapolated to higher redshifts constrained by the cosmic evolution of the SFR density, derived from deep ALMA imaging of the {\it Hubble Ultra Deep Field} in \citet{Dunlop_2017}.
Further details will be presented by Nava-Moreno et al. (in prep). This simulation reproduces the observed properties of the DSFG population with good accuracy and preserves the clustering information from the underlying dark-matter halo simulations.

We construct mock observations of red-\textit{Herschel} sources at 250, 350, and 500\,$\mu$m. To achieve this, we focus on sources with flux densities above the 250 SPIRE/\textit{Herschel} confusion limit \citep[$S_{250\mathrm{\mu m}} \sim 5.8$\,mJy,][]{Nguyen_2010} and coadd the flux density of all sources within a $\sim$9\,arcseconds radius (i.e., angular resolution at 250\,$\mu$m). The resulting flux density is treated as a single-dish measurement at 250\,$\mu$m. Similarly, simulated observations at 350 and 500\,$\mu$m are performed by centering on the same sources and summing the flux densities within a beam-size of 24 and 36\,arcseconds, which approximately corresponds to the \textit{Herschel}/SPIRE angular resolution at these wavelengths. 
We select all sources that meet our criteria: $S_{250 \mu \mathrm{m}} < S_{350 \mu \mathrm{m}} < S_{500 \mu \mathrm{m}}$ and $S_{500 \mu \mathrm{m}}>40$\,mJy, the adopted detection limit for the \textit{H}-ATLAS catalogs. 
These simulated red-\textit{Herschel} sources are used to construct a mock catalog to serve as input for the simulated ALMA observations at 1.3\,mm. For the Nava-Moreno et al. (in prep.) catalogue, the available 1.4\,mm flux densities were scaled to 1.3\,mm using the Rayleigh-Jeans approximation with an emissivity index of $\beta$=1.8. Our simulated ALMA observations are performed over areas of $r=16$\,arcsecond, the detection radius at 1.3\,mm, centered on each simulated red-\textit{Herschel} source to identify all DSFGs within a single ALMA pointing. We select all galaxies with flux densities above the ALMA detection limit of the HARPAS catalogue, corresponding to $S_{1.3,\mathrm{mm}} \sim 0.4$\,mJy (signal-to-noise ratio, SNR $\gtrsim$ 5). Based on the HARPAS criteria for each subsample, we classify all fields according to the number and distance between the detected 1.3\,mm sources into S-Fields, PLUM-Fields, M-Fields and fields with no detections.

We produced 368 pointings, of which 349 contain at least one ALMA-like source with $S_{1.3\,\mathrm{mm}}>$0.4\,mJy (the faintest ALMA source in the HARPAS catalogue). After that, we incorporate the completeness of the HARPAS catalogue (Appendix. \ref{apex:complet}). This results in 286 simulated pointings with ALMA-like detections from which, 185 pointings are classified as S-Fields, 4 as PLUM-Fields, and 97 as M-Fields. The maximum number of sources in a single field is four, consistent with what is found in the HARPAS catalogue. The predicted multiplicity fraction is $\sim$34\,per\,cent, which is approximately 1.7 times the multiplicity in HARPAS, suggesting that the simulation predicts a higher occurrence of multiplicity than that observed in our sample. This discrepancy may result from simplifications in the simulation, such as not modeling the noise properties of the maps and instead applying a flux density threshold, although we note that completeness corrections have been taken into account.   

\begin{figure}
    \centering
    \includegraphics[width=\linewidth]{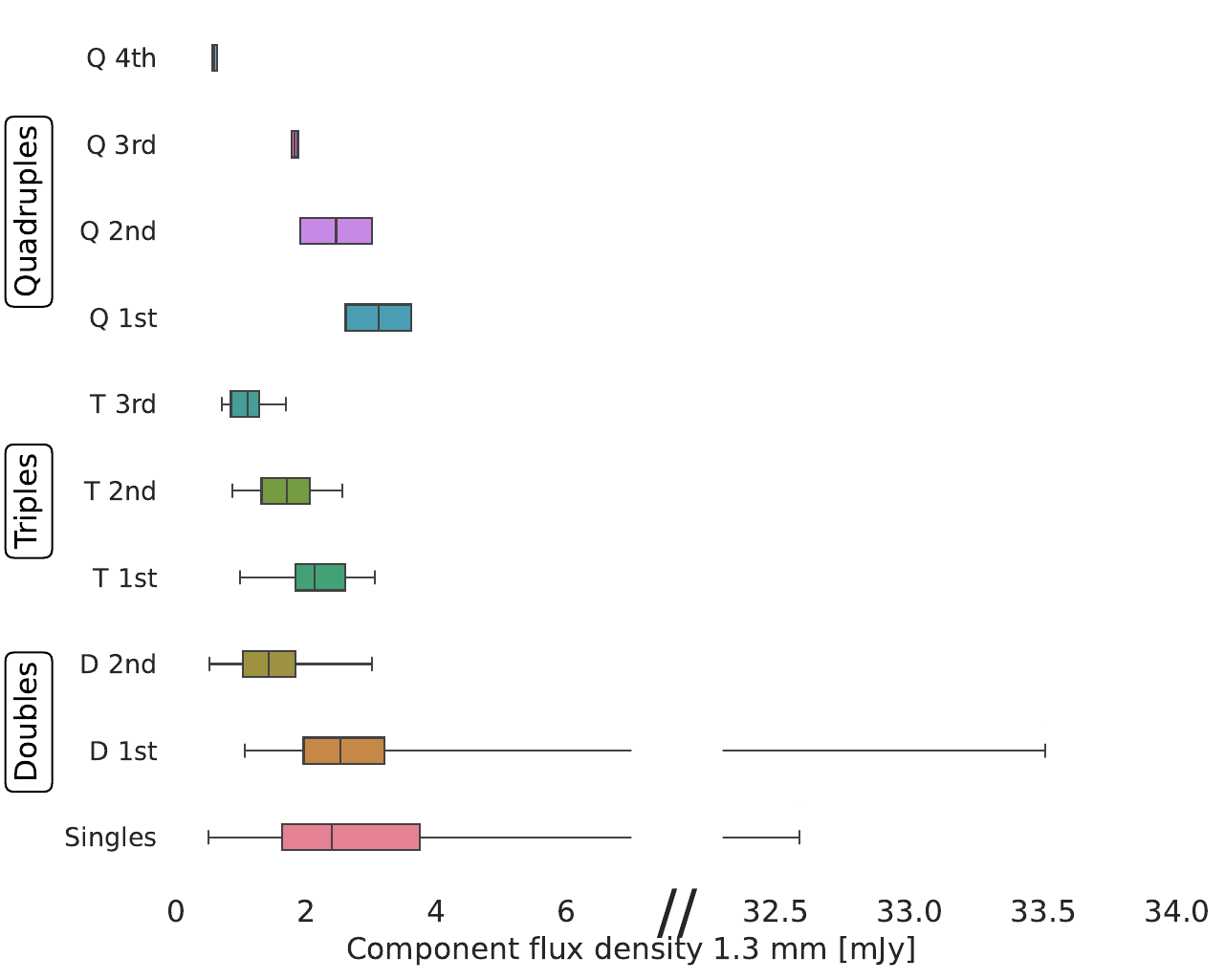}
    \caption{Flux densities from simulated ALMA pointings over 100\,square\,degrees (Nava-Moreno et al, in prep.), analogous to the HARPAS sample shown in Fig. \ref{fig:all_fluxes}. The vertical line in the box represents the 50\,per\,cent of the distribution, the limits of the box correspond to the 25th and 75th percentiles, and the lines extending from each box capture the range of the remaining data.
    }
    \label{fig:sim_ara_fluxes}
\end{figure}

Figure \ref{fig:sim_ara_fluxes} shows the 1.3 mm flux densities of the galaxies associated to simulated red-\textit{Herschel} sources. The median 1.3 mm flux density for the observed S-Fields is 3.1 mJy (68\,per\,cent range: 1.9-5.1 mJy), while the simulation yields 2.4 mJy (68\,per\,cent range: 1.4-5.8 mJy). Despite minor differences, the similar confidence intervals indicate that the simulation reproduces the observed flux density distribution properly. For the sub-samples of doubles, triples and quadruples, the brightest source contributes with a median of 64, 43, and 39\,per\,cent of the total flux density in each field, respectively. The median flux density ratios between the brightest and the fainter components of the systems are: 1.8 for doubles, 1.2 and 1.9 for triples, and 1.3, 1.7, and 5.3 for quadruples. These ratios are similar to those measured in the HARPAS sample (1.8 for doubles, 1.7 and 2.6 for triples, and 1.3, 2.4, and 4.0 for quadruples). Moreover, the simulated sources are generally as bright as the observed ones; however, they reach values of around 35\,mJy at 1.3\,mm (uncorrected for gravitational amplification), which are not observed in the M-Fields.
The median separation between sources in the simulations are 13.9, 11.3, and 13.7\,arcseconds for the three sub-classifications (i.e., doubles, triples, and quadruples), slightly higher ($\sim$ 3\,arcseconds) than those measured in the observed sample. 

We investigate whether the ALMA-like sources in each pointing are physically associated or merely projections along the line of sight. To identify associations, we adopt a redshift difference criterion of $\Delta z<0.01$, a threshold used in the literature \citep[e.g.,][]{Scudder_2018}, which is equivalent to a comoving radial distance of 10\,Mpc at $z\sim3$. Applying this criterion and considering all the ALMA-like sources within a specific field, our simulation predicts that 32\,per\,cent of doubles and 8\,per\,cent of triples are physically associated, while none of the quadruples meets the association criterion, indicating that none of the four sources in a field form a physically bound system.

\par Some observational results align with the findings of our mock sample. For instance, \citet{Simpson_2020} studied a sample of the 180 brightest sources in the SCUBA-2 850\,$\mu$m map of the COSMOS field with ALMA 870\,$\mu$m follow-up observations and reported that approximately 30\,per\,cent of those sources are physically associated when both observations and simulations are considered. This result is similar to that found by \citet{Scudder_2018}, who studied a sample of 360 \textit{Herschel}-detected sources at 250\,$\mu$m in the COSMOS field using 3.6 and 24\,$\mu$m \textit{Spitzer} catalogs and reported that 72\,per\,cent of the doubles in their sample exhibit no overlap in their redshift distributions. Furthermore, \citet{Stach_2018} analysed the initial AS2UDS results, which is an 870~$\mu$m ALMA continuum survey of 716 submillimeter sources from the SCUBA-2 Cosmology Legacy Survey (S2CLS) in the UKIDSS/UDS field, and reported that among 46 single-dish detections resolved into multiple ALMA sources with photometric redshifts, at least 30 per cent correspond to physically associated SMG pairs. Similarly, \citet{Wardlow_2018} examined six single-dish 870 $\mu$m sources in the Extended Chandra Deep Field-South and UKIDSS Ultra-Deep Survey regions, finding that $\sim$ 17 per cent of the multiple SMG components within these blended systems are physically related. These percentages are consistent with those derived from our mock sample, and indicate that the majority of the multiple systems are not necessarily physically associated.

\par Moreover, we estimate the probability of at least two ALMA-like sources in a field being physically associated, finding $\sim$67\,per\,cent for triples and 100\,per\,cent for quadruples. These results suggest that, even if not all ALMA-like sources within a field are physically associated, systems containing more than two sources still have a high probability that at least two are physically related. This implies that some of these multiple systems might still trace overdensities and could be used to identify galaxy proto-clusters. For example, one of the fields in our sample, classified as a triple (see Fig. \ref{fig:frac_t_q_1}), is the Distant Red Core (DRC) \citep{Oteo_2018}, named for its red colors in the SPIRE/\textit{Herschel} bands. This protocluster has been confirmed through follow-up studies to host at least ten DSFGs at $z=4.002$ over an area of 0.44 square arcminutes, with a collective star formation rate of $\sim$6,500\,M$_\odot$\,yr$^{-1}$ \citep{Oteo_2018}.

In summary, the simulation of ALMA-like sources presented here suggests that full physical associations in multiple systems are rare: only 32 per cent of doubles and 8 per cent of triples satisfy the strict redshift criterion, while no quadruples are fully bound. Nevertheless, the likelihood that at least two sources in a field are physically associated is higher, reaching $\sim$67 per cent for triples and 100 per cent for quadruples. Photometric-redshift analyses support this picture: subsets of sources in triples and quadruples often show compatible median redshifts (47 per cent and 67 per cent, respectively), whereas full association across all components remains uncommon, as confirmed by Kruskal–Wallis tests. Together, these results indicate that while entire systems rarely form fully bound structures, partial associations among subsets are frequent. This consistency between datasets suggests that photometric redshifts, despite their uncertainties, are useful for identifying candidate physically linked sources, while the strict redshift criterion provides a conservative lower limit on fully bound systems. Overall, M-Fields appear to host complex arrangements where subsets of sources may trace overdensities or proto-cluster regions, even if complete binding across all components is uncommon.

\par To confirm whether the systems in our sample are truly physically associated or merely projections, spectroscopic observations are necessary, as photometric redshifts have high uncertainties and provide only a preliminary estimate.

\subsection{Non-detections: What are they?}\label{sec:N-Fields}
In order to understand the nature of our non-detections, we begin by analyzing our simulated observations (described in section \ref{sec:sim}). We identified 19 simulated red-\textit{Herschel} sources that did not contain sources brighter than 0.4 mJy at 1.3\,mm in our simulated ALMA-like maps. These 19 fields contain between two and ten galaxies, with a median of 6. The sources in these fields have a median flux density of 0.1\,mJy, and approximately 73\,per\,cent of them lie at $z<2$. Additionally, we have 63 fields expected to be undetectable due to completeness, of which 23 are single sources and 40 are multiple systems at 1.3\,mm. In total, these 82 fields with non-detections represent 22.3\,per\,cent of the maps with simulated red-\textit{Herschel} sources, a fraction closely matching to that in our observations (21.7\,per\,cent). This consistency suggests that fields with no detections can be fully explained by intrinsically fainter sources falling close or below the ALMA detection threshold ($5\sigma$), where the survey suffers from a larger incompleteness (Fig. \ref{fig:completeness}). These simulated ALMA-like fields with no detections have a median redshift of 2.1 [1.1-2.8] and are dominated ($\sim 72$\,per\,cent) by multiple systems. Among these fields containing multiple sources, 21 fields ($\sim$ 36\,per\,cent) exhibit at least one physically associated pair of sources.

Furthermore, we investigate the 500\,$\mu$m properties of the simulated ALMA non-detections, which have a median flux density of $S_{500\mathrm{\mu m}}= 45.7$\,mJy, representing the faintest red-\textit{Herschel} sources in our sample. We resolve all the galaxies that contribute to the total flux density of each simulated red-\textit{Herschel} source, and identify those with $S_{500\mathrm{\mu m}}$ above the SPIRE confusion limit \citep[6.8\,mJy at 500\,$\mu$m;][]{Nguyen_2010}. We find at least one source above this limit in 59 of the ALMA fields (72\,per\,cent), while the remaining 23 (28\,per\,cent) contain multiple fainter components. ALMA fields with red-\textit{Herschel} sources above the confusion limit contribute to 52\,per\,cent of the total 500\,$\mu$m flux density, with the remaining 48\,per\,cent arising from multiple faint components that act as a flux boosting background. These 59 fields comprise 102 sources with a median 500\,$\mu$m flux density of $\sim$14\,mJy, and with only one source within one field meeting the red-\textit{Herschel} colour selection. Furthermore, these sources have a median redshift of $z \approx$ 2.4 (with $\sim$28\,per\,cent lying at $z < 2$) and therefore may correspond to a lower-$z$ population compared to those identified in the S-fields ($z \approx 2.78$ and less than 15\,per\,cent below $z = 2$, \citetalias{Quiros-Rojas2024}). This highlights how blending can result in the misidentification of red-\textit{Herschel} high-$z$ candidates.

To confirm if the non detections are explained by intrinsically fainter sources falling below the ALMA detection threshold, we analyze the observed 673 fields with no detections. Figure \ref{fig:number_sources} shows that the N-Fields correspond to the faintest sources in the sample at 500\,$\mu$m. Thus, to explain the absence of ALMA detections, we propose two scenarios: 1) the presence of multiple intrinsically faint sources with 1.3\,mm flux densities below our detection threshold,  as suggested by the simulation analysis presented above, and 2) false detections in SPIRE/\textit{Herschel} maps.

\begin{figure}
    \centering
    \includegraphics[width=\linewidth]{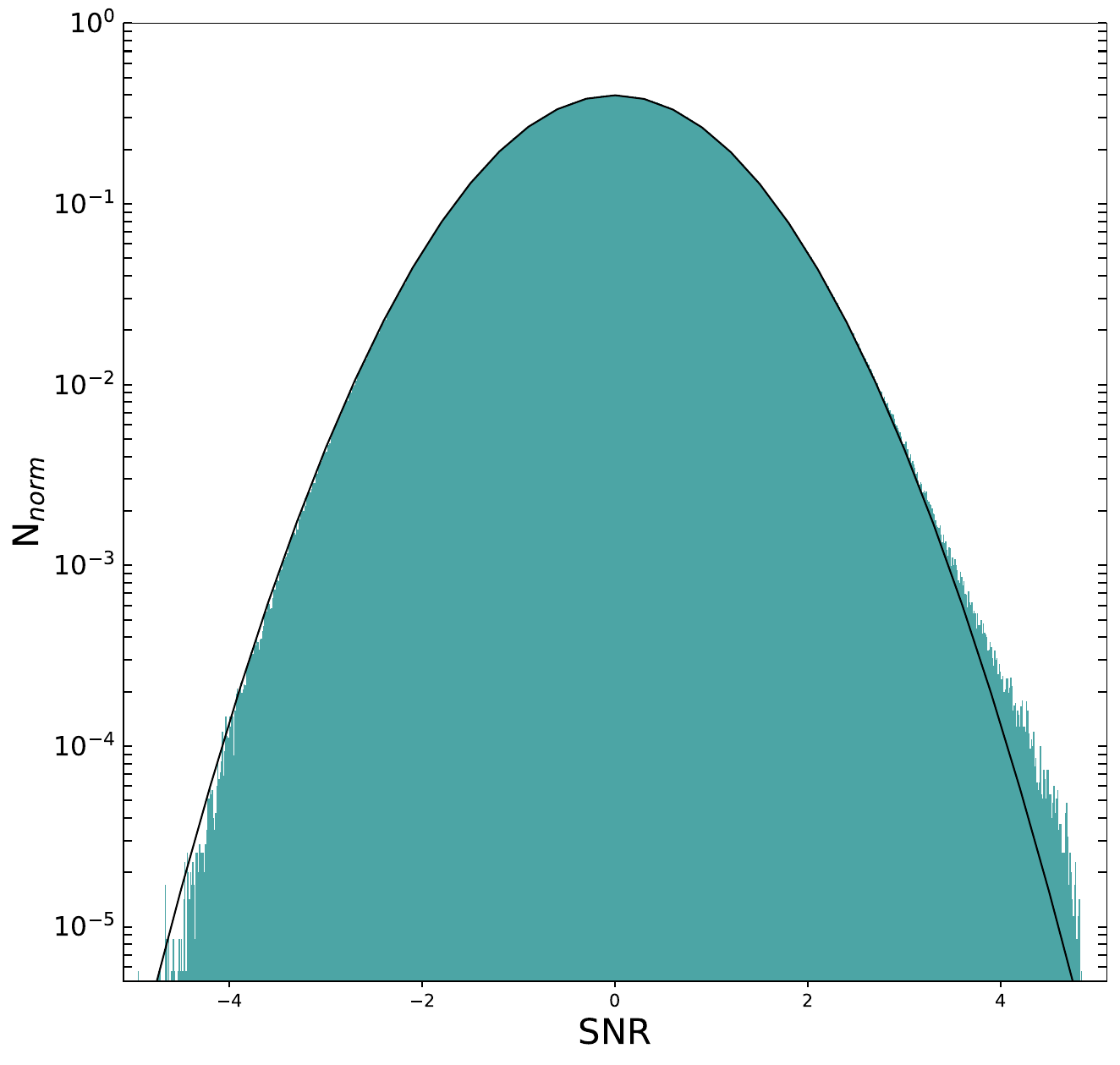}
    \caption{Number of pixels in the N-Fields as a function of signal-to-noise ratio (SNR).  The dark line represents a Gaussian fit to the data. An excess of approximately 4,500 pixels with SNR values between $\sim$3.5 and 5 suggests the presence of faint sources below our 5$\sigma$ ALMA detection limit.}
    \label{fig:SNR_NFields}
\end{figure}

To distinguish between these scenarios, we analyze the distribution of SNR in all pixel values at 1.3\,mm across N-Fields. If only noise is present, the distribution should closely follow a Gaussian function. However, our fit reveals an excess of approximately 4,500 pixels with SNR between $\sim$3.5 and 5, indicating the presence of faint sources below our ALMA detection limit  (see Fig. \ref{fig:SNR_NFields}).

To estimate the number of sources contributing to this excess, we calculate how the flux density of a single detection is distributed within an ALMA beam. Assuming a peak SNR of 4.25 (the centre of the observed excess), one detection would contribute signal to five pixels above the 3.5$\sigma$ threshold. Based on this, the observed excess corresponds to a lower limit of 900 sources. This implies that some N-Fields may host one or more sources at a $3.5<\mathrm{SNR}<5.0$ level. Under this assumption, the multiplicity fraction would rise to 40\,per\,cent.

In order to investigate whether most ALMA fields exhibit an excess of positive pixels, potentially caused by faint sources below the formal detection threshold, we compare the number of $3.5<\rm{SNR}<5.0$ detections in N-Fields and S-Fields, the dominant category in the HARPAS sample. After normalizing by the total number of maps in each category, we find an average of 3.5 faint sources in the S-fields compared to 3.8 in the N-fields. Although these values are not so different, the N-Fields exhibit a value that is $\sim$10\,per\,cent higher, which is in light with our current interpretation.\\

Thus, our results indicate that most of the non-detections in ALMA maps can be attributed to individual or multiple sources with fainter flux densities below our ALMA detection threshold.

\section{Summary and Conclusions}
We present an extension of the analysis of the HARPAS catalogue presented by \citet{Quiros-Rojas2024}. 
This study focuses on red-\textit{Herschel} sources that break into multiple components under higher angular resolution observations, with at least two sources separated by more than 3 arcseconds (M-Fields). This new category comprising 20\,per\,cent of the ALMA detections and includes 474 fields: 420 with two components, 51 with three, and 3 containing four components.

We investigate the feasibility of resolving this multiplicity with single-dish telescopes. For doubles, approximately 82, 40, and 24\,per\,cent would be resolved by TolTEC on LMT, NIKA2 on IRAM-30m, and SCUBA-2 on JCMT, respectively (with lower recovery fractions for triples and quadruples).

We also find that the combined effect of multiplicity and flux boosting introduces a factor of $\sim$2.6 overestimation in the number of $S_{500\mathrm{\mu m}} > 40$\,mJy red-\textit{Herschel} sources. This effect decreases with increasing $S_{500\mathrm{\mu m}}$ limit, with a $\sim$2.1 overestimation for sources with $S_{500\mathrm{\mu m}} > 80$\,mJy and a negligible impact at $S_{500\mathrm{\mu m}} > 120$\,mJy. However, we highlight that our results are based on a colour-selected sample, which may not fully reflect the broader population of \textit{Herschel} sources.

When analyzing the brightness distribution of our M-fields, we find that the median flux density of the brightest components in these multiple systems is comparable to that of the sources in S-Fields (i.e. fields with a single detection). The brightest source contributes, on average, 64, 48, and 42\,per\,cent of the total flux in doubles, triples, and quadruples, respectively. Moreover, the total flux density of the doubles is, on average, fainter than most of the triple or quadruple fields but consistent with the total flux density of their two brighter components.

M-Fields have a median $z_{\mathrm{phot}}$ of 3.0 [2.2–4.2], median $L_{\mathrm{IR}}$ of $6.0\,[3.1 - 11.1] \times 10^{12}\,\mathrm{L_{\odot}}$, a median SFR of $\sim 894\,[470-1640]\,\mathrm{M_{\odot}\, yr^{-1}}$, and a median molecular gas mass of $2.8\, [1.7-4.7] \times 10^{11}\, \mathrm{M_{\odot}}$. The values in brackets represent the range that encompasses 68\,per\,cent of the distribution (i.e, the 16th/84th percentiles). These properties are in agreement with those estimated for the sample of S-Fields in the HARPAS catalogue. Furthermore, we find that the SFRs of the sources identified in multiple systems, which may be influenced by their overdense environments, do not exceed 6,000 M$_{\odot}$ yr$^{-1}$, which is consistent with the upper limit found in \citetalias{Quiros-Rojas2024} for galaxies that are not affected by gravitational amplification.

To investigate whether doubles with evenly distributed flux densities between components could result from interactions, we analyzed the distances between components. However, no correlation was found between angular separation and flux density. Furthermore, we examined the possibility of physical association among sources within a field by analyzing if the medians of the sources are compatible or not with a common redshift distribution using non-parametric tests. For the doubles, 13\,per\,cent of the sources have median redshifts compatible with a common redshift distribution. For the triple and quadruple fields, the pairwise Mann–Whitney U tests indicate that 47 and 67\,per\,cent of the fields, respectively, contain at least one double with median redshifts consistent with sharing a common distribution. In contrast, the Kruskal–Wallis test reveals that only 2\,per\,cent of triple fields have median redshifts consistent with all components sharing a common origin, while none of the quadruple fields show such consistency between all four components.

When comparing the observed multiplicity properties with a mock redshift survey of the DSFG population \citep[][and Nava-Moreno et al., in prep.]{Nava_Moreno_2024}, we find that the simulation predicts a slightly higher occurrence of multiplicity than is observed in our data. Using the same simulation and applying a criterion of $\Delta z<0.01$ to identify physically associated galaxies we find that, when considering all sources in a given field, only 32\,per\,cent of doubles and 8\,per\,cent of triples are likely to be physically associated, while none of the quadruples are. However, our mock catalog suggests that approximately 67\,per\,cent of triples and 100\,per\,cent of quadruples contain at least two physically associated sources.  
The observations and the simulations thus suggest that the probability that all sources in these multiple source fields are physically associated is very low. Nevertheless, such fields can help trace overdensities, as the probability that subsets of sources share a similar redshift is higher.

Synthesizing the results obtained in this study with those from \citetalias{Quiros-Rojas2024}, and previous studies in the literature, we identify the following key implications for the interpretation of SMG populations:

High resolution ALMA observations show that most SMGs are individual within a region of 16.6\,arcsec (corresponding to $\sim$141\,kpc at $z \sim 2$). This suggests that large-scale interactions are unlikely to be the main driver of star formation in SMGs. Moreover, while closer interactions (of a few arcseconds) may contribute to the high star formation activity in some cases, they do not appear to be the dominant mechanism in this population either. The S-Fields are, by large, the dominat sources, even if considering only the brightest source of the system. Therefore, the main processes enhancing the SFR in these galaxies should occur on scales smaller than 1\,arcsec (corresponding to $\sim$8.5\,kpc at $z \sim 2$). Such processes might include coalescent mergers, disk instabilities, or turbulence (see also \citetalias{Quiros-Rojas2024}).

We also examine the fields with no ALMA detections in our sample and in our simulated ALMA-like fields with no detections, finding that most can be explained by individual or multiple sources with flux densities near or below the detection threshold ($5\sigma$). Taking these fields into account, the multiplicity fraction in our sample could reach up to $\sim$40\,per\,cent. These sources are, however, fainter than the M-fields, and based on the simulations, they seem to lie at slightly lower redshifts (with a median redshift of 2.1 [1.1–2.8]).

Both observations and simulations indicate that the probability that all sources in the multiple systems are physically associated is very low.  Nevertheless, some of these fields can still trace overdensities since some subsets of sources might still have similar redshift.

This study provides a catalog that facilitates further research into this extreme population of DSFGs, aiming to deepen our understanding of the dominant physical processes driving their evolution and to identify potential overdensities for follow-up observations.

\section*{acknowledgements}
We thank to the referee for insightful comments that have improved the manuscript. MQR would like to thank the La Secretaría de Ciencia, Humanidades, Tecnología e Innovación (SECIHTI) for her PhD grant. This work has been supported by the SECIHTI through projects A1-S-45680 and CB 2016 - 281948. JAZ acknowledge funding from JSPS KAKENHI grant number KG23K13150. IA acknowledge funding by Milgatz Grant ATR2024-154316 funded by MICIU/AEI/10.13039/501100011033, PID2022-136598NB-C33 by MCIN/AEI/10.13039/501100011033 and by ``ERDF A way of making Europe”.
The \textit{Herschel}-ATLAS is a project from \textit{Herschel}, which is an ESA space observatory with science instruments provided by European-led Principal Investigator consortia and with important participation from NASA.
This paper use the following ALMA projects: ADS/JAO.ALMA\#2016.1.00087.S,  ADS/JAO.ALMA\#2016.1.00139.S, ADS/JAO.ALMA\#2017.1.00510.S, ADS/JAO.ALMA\#2018.1.00489.S, ADS/JAO.ALMA\#2018.1.00526.S, ADS/JAO.ALMA\#2021.1.01628.S and ADSJAO.ALMA\#2022.1.00432.S. ALMA is a partnership of ESO (representing its member states), NSF (USA), and NINS (Japan), together with NRC (Canada), MOST and ASIAA (Taiwan), and KASI (Republic of Korea), in cooperation with the Republic of Chile. The Joint ALMA Observatory is operated
by ESO, AUI/NRAO, and NAOJ. Funding for the Sloan Digital Sky Survey V has been provided by the Alfred P. Sloan Foundation, the Heising-Simons Foundation, the National Science Foundation, and the Participating Institutions. SDSS acknowledges support and resources from the Center for High-Performance Computing at the University of Utah. SDSS telescopes are located at Apache Point Observatory, funded by the Astrophysical Research Consortium and operated by New Mexico State University, and at Las Campanas Observatory, operated by the Carnegie Institution for Science. The SDSS web site is \url{www.sdss.org}.
SDSS is managed by the Astrophysical Research Consortium for the Participating Institutions of the SDSS Collaboration, including Caltech, the Carnegie Institution for Science, Chilean National Time Allocation Committee (CNTAC) ratified researchers, The Flatiron Institute, the Gotham Participation Group, Harvard University, Heidelberg University, The Johns Hopkins University, L’Ecole polytechnique fédérale de Lausanne (EPFL), Leibniz-Institut für Astrophysik Potsdam (AIP), Max-Planck-Institut für Astronomie (MPIA Heidelberg), Max-Planck-Institut für Extraterrestrische Physik (MPE), Nanjing University, National Astronomical Observatories of China (NAOC), New Mexico State University, The Ohio State University, Pennsylvania State University, Smithsonian Astrophysical Observatory, Space Telescope Science Institute (STScI), the Stellar Astrophysics Participation Group, Universidad Nacional Autónoma de México, University of Arizona, University of Colorado Boulder, University of Illinois at Urbana-Champaign, University of Toronto, University of Utah, University of Virginia, Yale University, and Yunnan University.

%

%
\section*{Data Availability}
Catalogues are available at \url{https://mnemosyne.inaoep.mx/index.php/s/BgSmIkbTsryVma9} and included as supplementary data at MNRAS online. The \textit{Herschel} Astrophysical Terahertz Large Area Survey (\textit{H}-ATLAS) catalogs and maps are available in: \url{https://www.h-atlas.org/public-data}. The ALMA data band 6 is available in 
\url{ https://data.nrao.edu/portal/} for the MS files and in \url{https://almascience.eso.org/aq/} for the maps.



\bibliographystyle{mnras}
\bibliography{aa} 

\begin{thebibliography}{}
\makeatletter
\relax
\def\mn@urlcharsother{\let\do\@makeother \do\$\do\&\do\#\do\^\do\_\do\%\do\~}
\def\mn@doi{\begingroup\mn@urlcharsother \@ifnextchar [ {\mn@doi@} {\mn@doi@[]}}
\def\mn@doi@[#1]#2{\def\@tempa{#1}\ifx\@tempa\@empty \href {http://dx.doi.org/#2} {doi:#2}\else \href {http://dx.doi.org/#2} {#1}\fi \endgroup}
\def\mn@eprint#1#2{\mn@eprint@#1:#2::\@nil}
\def\mn@eprint@arXiv#1{\href {http://arxiv.org/abs/#1} {{\tt arXiv:#1}}}
\def\mn@eprint@dblp#1{\href {http://dblp.uni-trier.de/rec/bibtex/#1.xml} {dblp:#1}}
\def\mn@eprint@#1:#2:#3:#4\@nil{\def\@tempa {#1}\def\@tempb {#2}\def\@tempc {#3}\ifx \@tempc \@empty \let \@tempc \@tempb \let \@tempb \@tempa \fi \ifx \@tempb \@empty \def\@tempb {arXiv}\fi \@ifundefined {mn@eprint@\@tempb}{\@tempb:\@tempc}{\expandafter \expandafter \csname mn@eprint@\@tempb\endcsname \expandafter{\@tempc}}}

\bibitem[\protect\citeauthoryear{Allen, Evrard  \& Mantz}{Allen et~al.}{2011}]{Allen_2011}
Allen S.~W.,  Evrard A.~E.,   Mantz A.~B.,  2011, \mn@doi [Annual Review of Astronomy and Astrophysics] {https://doi.org/10.1146/annurev-astro-081710-102514}, 49, 409

\bibitem[\protect\citeauthoryear{Aravena et~al.,}{Aravena et~al.}{2016}]{Aravena_2016}
Aravena M.,  et~al., 2016, \mn@doi [Monthly Notices of the Royal Astronomical Society] {10.1093/mnras/stw275}, 457, 4406

\bibitem[\protect\citeauthoryear{Aretxaga, Hughes  \& Dunlop}{Aretxaga et~al.}{2005}]{Aretxaga_2005}
Aretxaga I.,  Hughes D.~H.,   Dunlop J.~S.,  2005, \mn@doi [Monthly Notices of the Royal Astronomical Society] {10.1111/j.1365-2966.2005.08733.x}, 358, 1240

\bibitem[\protect\citeauthoryear{Aretxaga et~al.,}{Aretxaga et~al.}{2007}]{Aretxaga_2007}
Aretxaga I.,  et~al., 2007, \mn@doi [Monthly Notices of the Royal Astronomical Society] {10.1111/j.1365-2966.2007.12036.x}, 379, 1571

\bibitem[\protect\citeauthoryear{Baugh, Lacey, Frenk, Granato, Silva, Bressan, Benson  \& Cole}{Baugh et~al.}{2005}]{Baugh_2005}
Baugh C.~M.,  Lacey C.~G.,  Frenk C.~S.,  Granato G.~L.,  Silva L.,  Bressan A.,  Benson A.~J.,   Cole S.,  2005, \mn@doi [Monthly Notices of the Royal Astronomical Society] {10.1111/j.1365-2966.2004.08553.x}, 356, 1191

\bibitem[\protect\citeauthoryear{Bendo et~al.,}{Bendo et~al.}{2022}]{Bendo}
Bendo G.~J.,  et~al., 2022, \mn@doi [Monthly Notices of the Royal Astronomical Society] {10.1093/mnras/stac3771}, 522, 2995

\bibitem[\protect\citeauthoryear{{Berta} et~al.,}{{Berta} et~al.}{2023}]{Berta_2023}
{Berta} et~al., 2023, \mn@doi [A\&A] {10.1051/0004-6361/202346803}, 678, A28

\bibitem[\protect\citeauthoryear{Birkin et~al.,}{Birkin et~al.}{2020}]{Birkin_2020}
Birkin J.~E.,  et~al., 2020, \mn@doi [Monthly Notices of the Royal Astronomical Society] {10.1093/mnras/staa3862}, 501, 3926

\bibitem[\protect\citeauthoryear{{Blain}, {Smail}, {Ivison}, {Kneib}  \& {Frayer}}{{Blain} et~al.}{2002}]{Blain2002}
{Blain} A.~W.,  {Smail} I.,  {Ivison} R.~J.,  {Kneib} J.~P.,   {Frayer} D.~T.,  2002, \mn@doi [\physrep] {10.1016/S0370-1573(02)00134-5}, \href {https://ui.adsabs.harvard.edu/abs/2002PhR...369..111B} {369, 111}

\bibitem[\protect\citeauthoryear{Bothwell et~al.,}{Bothwell et~al.}{2013}]{bothwell}
Bothwell M.~S.,  et~al., 2013, \mn@doi [Monthly Notices of the Royal Astronomical Society] {10.1093/mnras/sts562}, 429, 3047

\bibitem[\protect\citeauthoryear{Cairns et~al.,}{Cairns et~al.}{2022}]{Cairns_2022}
Cairns J.,  et~al., 2022, \mn@doi [Monthly Notices of the Royal Astronomical Society] {10.1093/mnras/stac3486}, 519, 709

\bibitem[\protect\citeauthoryear{Casey}{Casey}{2020}]{Casey_2020}
Casey C.~M.,  2020, \mn@doi [The Astrophysical Journal] {10.3847/1538-4357/aba528}, 900, 68

\bibitem[\protect\citeauthoryear{{Casey}, {Narayanan}  \& {Cooray}}{{Casey} et~al.}{2014}]{Casey2014}
{Casey} C.~M.,  {Narayanan} D.,   {Cooray} A.,  2014, \mn@doi [\physrep] {10.1016/j.physrep.2014.02.009}, \href {https://ui.adsabs.harvard.edu/abs/2014PhR...541...45C} {541, 45}

\bibitem[\protect\citeauthoryear{Casey et~al.,}{Casey et~al.}{2018}]{Casey_2018}
Casey C.~M.,  et~al., 2018, \mn@doi [The Astrophysical Journal] {10.3847/1538-4357/aac82d}, 862, 77

\bibitem[\protect\citeauthoryear{Chapman, Blain, Smail  \& Ivison}{Chapman et~al.}{2005}]{Chapman_2005}
Chapman S.~C.,  Blain A.~W.,  Smail I.,   Ivison R.~J.,  2005, \mn@doi [The Astrophysical Journal] {10.1086/428082}, 622, 772

\bibitem[\protect\citeauthoryear{Chen, Cowie, Barger, Casey, Lee, Sanders, Wang  \& Williams}{Chen et~al.}{2013}]{scuba-2}
Chen C.-C.,  Cowie L.~L.,  Barger A.~J.,  Casey C.~M.,  Lee N.,  Sanders D.~B.,  Wang W.-H.,   Williams J.~P.,  2013, \mn@doi [The Astrophysical Journal] {10.1088/0004-637X/776/2/131}, 776, 131

\bibitem[\protect\citeauthoryear{Cooray \& Sheth}{Cooray \& Sheth}{2002}]{COORAY20021}
Cooray A.,  Sheth R.,  2002, \mn@doi [Physics Reports] {https://doi.org/10.1016/S0370-1573(02)00276-4}, 372, 1

\bibitem[\protect\citeauthoryear{Cornish et~al.,}{Cornish et~al.}{2024}]{Cornish_2024}
Cornish T.~M.,  et~al., 2024, \mn@doi [Monthly Notices of the Royal Astronomical Society] {10.1093/mnras/stae1868}, 533, 2399

\bibitem[\protect\citeauthoryear{Cox et~al.,}{Cox et~al.}{2011}]{Cox_2011}
Cox P.,  et~al., 2011, \mn@doi [The Astrophysical Journal] {10.1088/0004-637X/740/2/63}, 740, 63

\bibitem[\protect\citeauthoryear{{Cox} et~al.,}{{Cox} et~al.}{2023}]{Cox_2024}
{Cox} et~al., 2023, \mn@doi [A\&A] {10.1051/0004-6361/202346801}, 678, A26

\bibitem[\protect\citeauthoryear{Dudzevi{\v{c}}i{\=u}t{\.e} et~al.,}{Dudzevi{\v{c}}i{\=u}t{\.e} et~al.}{2020}]{dudzevivciute2020}
Dudzevi{\v{c}}i{\=u}t{\.e} U.,  et~al., 2020, Monthly Notices of the Royal Astronomical Society, 494, 3828

\bibitem[\protect\citeauthoryear{Dudzevičiūtė et~al.,}{Dudzevičiūtė et~al.}{2020}]{Dudze_2020}
Dudzevičiūtė U.,  et~al., 2020, \mn@doi [Monthly Notices of the Royal Astronomical Society] {10.1093/mnras/staa769}, 494, 3828

\bibitem[\protect\citeauthoryear{Dunlop et~al.,}{Dunlop et~al.}{2016}]{Dunlop_2017}
Dunlop J.~S.,  et~al., 2016, \mn@doi [Monthly Notices of the Royal Astronomical Society] {10.1093/mnras/stw3088}, 466, 861

\bibitem[\protect\citeauthoryear{Eales et~al.,}{Eales et~al.}{2010}]{Eales_2010}
Eales S.,  et~al., 2010, \mn@doi [Publications of the Astronomical Society of the Pacific] {10.1086/653086}, 122, 499

\bibitem[\protect\citeauthoryear{{Franco} et~al.,}{{Franco} et~al.}{2018}]{Franco_2018}
{Franco} et~al., 2018, \mn@doi [A\&A] {10.1051/0004-6361/201832928}, 620, A152

\bibitem[\protect\citeauthoryear{{Golec, J.} \& {The ToITEC Collaboration}}{{Golec, J.} \& {The ToITEC Collaboration}}{2024}]{toltec}
{Golec, J.} {The ToITEC Collaboration} 2024, \mn@doi [EPJ Web Conf.] {10.1051/epjconf/202429300022}, 293, 00022

\bibitem[\protect\citeauthoryear{Greenslade, Clements, Petitpas, Asboth, Conley, Pérez-Fournon  \& Riechers}{Greenslade et~al.}{2020}]{Greenslade_2020}
Greenslade J.,  Clements D.~L.,  Petitpas G.,  Asboth V.,  Conley A.,  Pérez-Fournon I.,   Riechers D.,  2020, \mn@doi [Monthly Notices of the Royal Astronomical Society] {10.1093/mnras/staa1637}, 496, 2315

\bibitem[\protect\citeauthoryear{{Griffin, M. J.} et~al.,}{{Griffin, M. J.} et~al.}{2010}]{Griffin_2010}
{Griffin, M. J.} et~al., 2010, \mn@doi [A&A] {10.1051/0004-6361/201014519}, 518, L3

\bibitem[\protect\citeauthoryear{{Gururajan} et~al.,}{{Gururajan} et~al.}{2022}]{Gururajan}
{Gururajan} et~al., 2022, \mn@doi [A\&A] {10.1051/0004-6361/202142172}, 663, A22

\bibitem[\protect\citeauthoryear{Hatsukade et~al.,}{Hatsukade et~al.}{2018}]{HATSUKADE_2018}
Hatsukade B.,  et~al., 2018, \mn@doi [Publications of the Astronomical Society of Japan] {10.1093/pasj/psy104}, 70, 105

\bibitem[\protect\citeauthoryear{Hayward, Kereš, Jonsson, Narayanan, Cox  \& Hernquist}{Hayward et~al.}{2011}]{Hayward_2011}
Hayward C.~C.,  Kereš D.,  Jonsson P.,  Narayanan D.,  Cox T.~J.,   Hernquist L.,  2011, \mn@doi [The Astrophysical Journal] {10.1088/0004-637X/743/2/159}, 743, 159

\bibitem[\protect\citeauthoryear{Hayward, Behroozi, Somerville, Primack, Moreno  \& Wechsler}{Hayward et~al.}{2013}]{Hayward_2013}
Hayward C.~C.,  Behroozi P.~S.,  Somerville R.~S.,  Primack J.~R.,  Moreno J.,   Wechsler R.~H.,  2013, \mn@doi [Monthly Notices of the Royal Astronomical Society] {10.1093/mnras/stt1202}, 434, 2572

\bibitem[\protect\citeauthoryear{Herwig, Battaia, Chen, Obreja, Nowotka, Remus  \& Yajima}{Herwig et~al.}{2025}]{herwig_2025}
Herwig E.,  Battaia F.~A.,  Chen C.-C.,  Obreja A.,  Nowotka M.,  Remus R.-S.,   Yajima H.,  2025, Submillimeter galaxy overdensities around physically associated quasar pairs (\mn@eprint {arXiv} {2506.11193}), \url {https://arxiv.org/abs/2506.11193}

\bibitem[\protect\citeauthoryear{Hodge et~al.,}{Hodge et~al.}{2013}]{Hodge_2013}
Hodge J.~A.,  et~al., 2013, \mn@doi [The Astrophysical Journal] {10.1088/0004-637X/768/1/91}, 768, 91

\bibitem[\protect\citeauthoryear{{Hughes} et~al.,}{{Hughes} et~al.}{1998}]{Hughes1998}
{Hughes} D.~H.,  et~al., 1998, \mn@doi [\nat] {10.1038/28328}, \href {https://ui.adsabs.harvard.edu/abs/1998Natur.394..241H} {394, 241}

\bibitem[\protect\citeauthoryear{Hughes et~al.,}{Hughes et~al.}{2002}]{Hughes2002}
Hughes D.~H.,  et~al., 2002, \mn@doi [Monthly Notices of the Royal Astronomical Society] {10.1046/j.1365-8711.2002.05670.x}, 335, 871

\bibitem[\protect\citeauthoryear{Hughes et~al.,}{Hughes et~al.}{2020}]{LMT}
Hughes D.~H.,  et~al., 2020, in Marshall H.~K.,  Spyromilio J.,   Usuda T.,  eds,  Vol. 11445, Ground-based and Airborne Telescopes VIII. SPIE, p. 1144522, \mn@doi{10.1117/12.2561893}, \url {https://doi.org/10.1117/12.2561893}

\bibitem[\protect\citeauthoryear{Hung et~al.,}{Hung et~al.}{2016}]{Hung_2016}
Hung C.-L.,  et~al., 2016, \mn@doi [The Astrophysical Journal] {10.3847/0004-637X/826/2/130}, 826, 130

\bibitem[\protect\citeauthoryear{Hurley et~al.,}{Hurley et~al.}{2016}]{XID+}
Hurley P.~D.,  et~al., 2016, \mn@doi [Monthly Notices of the Royal Astronomical Society] {10.1093/mnras/stw2375}, 464, 885

\bibitem[\protect\citeauthoryear{{Ikarashi}, {Ivison, R. J.}, {Cowley, W. I.}  \& {Kohno, K.}}{{Ikarashi} et~al.}{2022}]{Ikarashi_2022}
{Ikarashi} {Ivison, R. J.} {Cowley, W. I.}  {Kohno, K.} 2022, \mn@doi [A\&A] {10.1051/0004-6361/202141196}, 659, A154

\bibitem[\protect\citeauthoryear{Kennicutt \& Evans}{Kennicutt \& Evans}{2012}]{Kennicutt_2012}
Kennicutt R.~C.,  Evans N.~J.,  2012, \mn@doi [Annual Review of Astronomy and Astrophysics] {10.1146/annurev-astro-081811-125610}, 50, 531

\bibitem[\protect\citeauthoryear{Kravtsov \& Borgani}{Kravtsov \& Borgani}{2012}]{Kractsov_2012}
Kravtsov A.~V.,  Borgani S.,  2012, \mn@doi [Annual Review of Astronomy and Astrophysics] {https://doi.org/10.1146/annurev-astro-081811-125502}, 50, 353

\bibitem[\protect\citeauthoryear{Kroupa}{Kroupa}{2001}]{Kroupa_2001}
Kroupa P.,  2001, \mn@doi [Monthly Notices of the Royal Astronomical Society] {10.1046/j.1365-8711.2001.04022.x}, 322, 231

\bibitem[\protect\citeauthoryear{Lacey, Baugh, Frenk, Benson, Orsi, Silva, Granato  \& Bressan}{Lacey et~al.}{2010}]{Lacey_2010}
Lacey C.~G.,  Baugh C.~M.,  Frenk C.~S.,  Benson A.~J.,  Orsi A.,  Silva L.,  Granato G.~L.,   Bressan A.,  2010, \mn@doi [Monthly Notices of the Royal Astronomical Society] {10.1111/j.1365-2966.2010.16463.x}, 405, 2

\bibitem[\protect\citeauthoryear{Lacey et~al.,}{Lacey et~al.}{2016}]{Lacey_2016}
Lacey C.~G.,  et~al., 2016, \mn@doi [Monthly Notices of the Royal Astronomical Society] {10.1093/mnras/stw1888}, 462, 3854

\bibitem[\protect\citeauthoryear{Ma et~al.,}{Ma et~al.}{2019}]{Ma_2019}
Ma J.,  et~al., 2019, \mn@doi [The Astrophysical Journal Supplement Series] {10.3847/1538-4365/ab4194}, 244, 30

\bibitem[\protect\citeauthoryear{Maddox et~al.,}{Maddox et~al.}{2018}]{Maddox_2018}
Maddox S.~J.,  et~al., 2018, \mn@doi [The Astrophysical Journal Supplement Series] {10.3847/1538-4365/aab8fc}, 236, 30

\bibitem[\protect\citeauthoryear{McKinney et~al.,}{McKinney et~al.}{2025}]{McKinney_2025}
McKinney J.,  et~al., 2025, \mn@doi [The Astrophysical Journal] {10.3847/1538-4357/ada357}, 979, 229

\bibitem[\protect\citeauthoryear{{Miller} et~al.,}{{Miller} et~al.}{2018}]{Miller_2017}
{Miller} T.~B.,  et~al., 2018, \mn@doi [\nat] {10.1038/s41586-018-0025-2}, \href {https://ui.adsabs.harvard.edu/abs/2018Natur.556..469M} {556, 469}

\bibitem[\protect\citeauthoryear{Montaña et~al.,}{Montaña et~al.}{2021}]{Montaña_2021}
Montaña A.,  et~al., 2021, \mn@doi [Monthly Notices of the Royal Astronomical Society] {10.1093/mnras/stab1649}, 505, 5260

\bibitem[\protect\citeauthoryear{Nava-Moreno, Montaña, Aretxaga, Rodríguez-Puebla, Avila-Reese  \& Peralta}{Nava-Moreno et~al.}{2024}]{Nava_Moreno_2024}
Nava-Moreno N.~A.,  Montaña A.,  Aretxaga I.,  Rodríguez-Puebla A.,  Avila-Reese V.,   Peralta E.,  2024, \mn@doi [Monthly Notices of the Royal Astronomical Society] {10.1093/mnras/stae1417}, 531, 4900

\bibitem[\protect\citeauthoryear{Negrello et~al.,}{Negrello et~al.}{2010}]{negrello_2010}
Negrello M.,  et~al., 2010, science, 330, 800

\bibitem[\protect\citeauthoryear{{Nguyen} et~al.,}{{Nguyen} et~al.}{2010}]{Nguyen_2010}
{Nguyen} et~al., 2010, \mn@doi [A&A] {10.1051/0004-6361/201014680}, 518, L5

\bibitem[\protect\citeauthoryear{Oliver et~al.,}{Oliver et~al.}{2012}]{Oliver_2012}
Oliver S.~J.,  et~al., 2012, \mn@doi [Monthly Notices of the Royal Astronomical Society] {10.1111/j.1365-2966.2012.20912.x}, 424, 1614

\bibitem[\protect\citeauthoryear{Oteo et~al.,}{Oteo et~al.}{2018}]{Oteo_2018}
Oteo I.,  et~al., 2018, \mn@doi [The Astrophysical Journal] {10.3847/1538-4357/aaa1f1}, 856, 72

\bibitem[\protect\citeauthoryear{Overzier}{Overzier}{2016}]{Overzier+2016}
Overzier R.~A.,  2016, \mn@doi [The Astronomy and Astrophysics Review] {10.1007/s00159-016-0100-3}, 24

\bibitem[\protect\citeauthoryear{Peck, Schinckel  \& team}{Peck et~al.}{2007}]{SMA}
Peck A.,  Schinckel A.,   team S.,  2007, in Lobanov A.~P.,  Zensus J.~A.,  Cesarsky C.,   Diamond P.~J.,  eds, Exploring the Cosmic Frontier. Springer Berlin Heidelberg, Berlin, Heidelberg, pp 49--50

\bibitem[\protect\citeauthoryear{{Perotto, L.} et~al.,}{{Perotto, L.} et~al.}{2020}]{nika2}
{Perotto, L.} et~al., 2020, \mn@doi [A\&A] {10.1051/0004-6361/201936220}, 637, A71

\bibitem[\protect\citeauthoryear{{Pilbratt, G. L.} et~al.,}{{Pilbratt, G. L.} et~al.}{2010}]{Pilbratt_2010}
{Pilbratt, G. L.} et~al., 2010, \mn@doi [A&A] {10.1051/0004-6361/201014759}, 518, L1

\bibitem[\protect\citeauthoryear{{Poglitsch, A.} et~al.,}{{Poglitsch, A.} et~al.}{2010}]{Poglitsch_2010}
{Poglitsch, A.} et~al., 2010, \mn@doi [A&A] {10.1051/0004-6361/201014535}, 518, L2

\bibitem[\protect\citeauthoryear{Pope \& Chary}{Pope \& Chary}{2010}]{Pope_2010}
Pope A.,  Chary R.-R.,  2010, \mn@doi [The Astrophysical Journal] {10.1088/2041-8205/715/2/l171}, 715, L171

\bibitem[\protect\citeauthoryear{Quirós-Rojas, Montaña, Zavala, Aretxaga  \& Hughes}{Quirós-Rojas et~al.}{2024}]{Quiros-Rojas2024}
Quirós-Rojas M.,  Montaña A.,  Zavala J.~A.,  Aretxaga I.,   Hughes D.~H.,  2024, \mn@doi [Monthly Notices of the Royal Astronomical Society] {10.1093/mnras/stae1974}, 533, 2966

\bibitem[\protect\citeauthoryear{Scoville et~al.,}{Scoville et~al.}{2016}]{Scoville_2016}
Scoville N.,  et~al., 2016, \mn@doi [The Astrophysical Journal] {10.3847/0004-637X/820/2/83}, 820, 83

\bibitem[\protect\citeauthoryear{Scudder, Oliver, Hurley, Wardlow, Wang  \& Farrah}{Scudder et~al.}{2018}]{Scudder_2018}
Scudder J.~M.,  Oliver S.,  Hurley P.~D.,  Wardlow J.~L.,  Wang L.,   Farrah D.,  2018, \mn@doi [Monthly Notices of the Royal Astronomical Society] {10.1093/mnras/sty2009}, 480, 4124

\bibitem[\protect\citeauthoryear{Simpson et~al.,}{Simpson et~al.}{2020}]{Simpson_2020}
Simpson J.~M.,  et~al., 2020, \mn@doi [Monthly Notices of the Royal Astronomical Society] {10.1093/mnras/staa1345}, 495, 3409

\bibitem[\protect\citeauthoryear{Smail, Ivison  \& Blain}{Smail et~al.}{1997}]{Smail_1997}
Smail I.,  Ivison R.~J.,   Blain A.~W.,  1997, \mn@doi [The Astrophysical Journal] {10.1086/311017}, 490, L5

\bibitem[\protect\citeauthoryear{Stach et~al.,}{Stach et~al.}{2018}]{Stach_2018}
Stach S.~M.,  et~al., 2018, \mn@doi [The Astrophysical Journal] {10.3847/1538-4357/aac5e5}, 860, 161

\bibitem[\protect\citeauthoryear{Valiante et~al.,}{Valiante et~al.}{2016}]{Valiante_2016}
Valiante E.,  et~al., 2016, \mn@doi [Monthly Notices of the Royal Astronomical Society] {10.1093/mnras/stw1806}, 462, 3146

\bibitem[\protect\citeauthoryear{Voit}{Voit}{2005}]{Voit_2005}
Voit G.~M.,  2005, \mn@doi [Rev. Mod. Phys.] {10.1103/RevModPhys.77.207}, 77, 207

\bibitem[\protect\citeauthoryear{{Wang} et~al.,}{{Wang} et~al.}{2024}]{Wang_2024}
{Wang} et~al., 2024, \mn@doi [A\&A] {10.1051/0004-6361/202349055}, 688, A20

\bibitem[\protect\citeauthoryear{Wardlow et~al.,}{Wardlow et~al.}{2018}]{Wardlow_2018}
Wardlow J.~L.,  et~al., 2018, \mn@doi [Monthly Notices of the Royal Astronomical Society] {10.1093/mnras/sty1526}, 479, 3879

\bibitem[\protect\citeauthoryear{Whitaker, Pope, Cybulski, Casey, Popping  \& Yun}{Whitaker et~al.}{2017}]{Whitaker_2017}
Whitaker K.~E.,  Pope A.,  Cybulski R.,  Casey C.~M.,  Popping G.,   Yun M.~S.,  2017, \mn@doi [The Astrophysical Journal] {10.3847/1538-4357/aa94ce}, 850, 208

\bibitem[\protect\citeauthoryear{Wootten \& Thompson}{Wootten \& Thompson}{2009}]{ALMA}
Wootten A.,  Thompson A.~R.,  2009, \mn@doi [Proceedings of the IEEE] {10.1109/JPROC.2009.2020572}, 97, 1463

\bibitem[\protect\citeauthoryear{{Yang} et~al.,}{{Yang} et~al.}{2017}]{yang_2017}
{Yang} et~al., 2017, \mn@doi [A\&A] {10.1051/0004-6361/201731391}, 608, A144

\bibitem[\protect\citeauthoryear{Yun et~al.,}{Yun et~al.}{2012}]{Yun_2012}
Yun M.~S.,  et~al., 2012, \mn@doi [Monthly Notices of the Royal Astronomical Society] {10.1111/j.1365-2966.2011.19898.x}, 420, 957

\bibitem[\protect\citeauthoryear{Zavala et~al.,}{Zavala et~al.}{2018}]{Zavala}
Zavala J.~A.,  et~al., 2018, Nature Astronomy, 2, 56

\makeatother
\end{thebibliography}




\appendix

\section{Completeness of the HARPAS sample}
\label{apex:complet}
\begin{figure}
    \centering
    \includegraphics[width=\hsize]{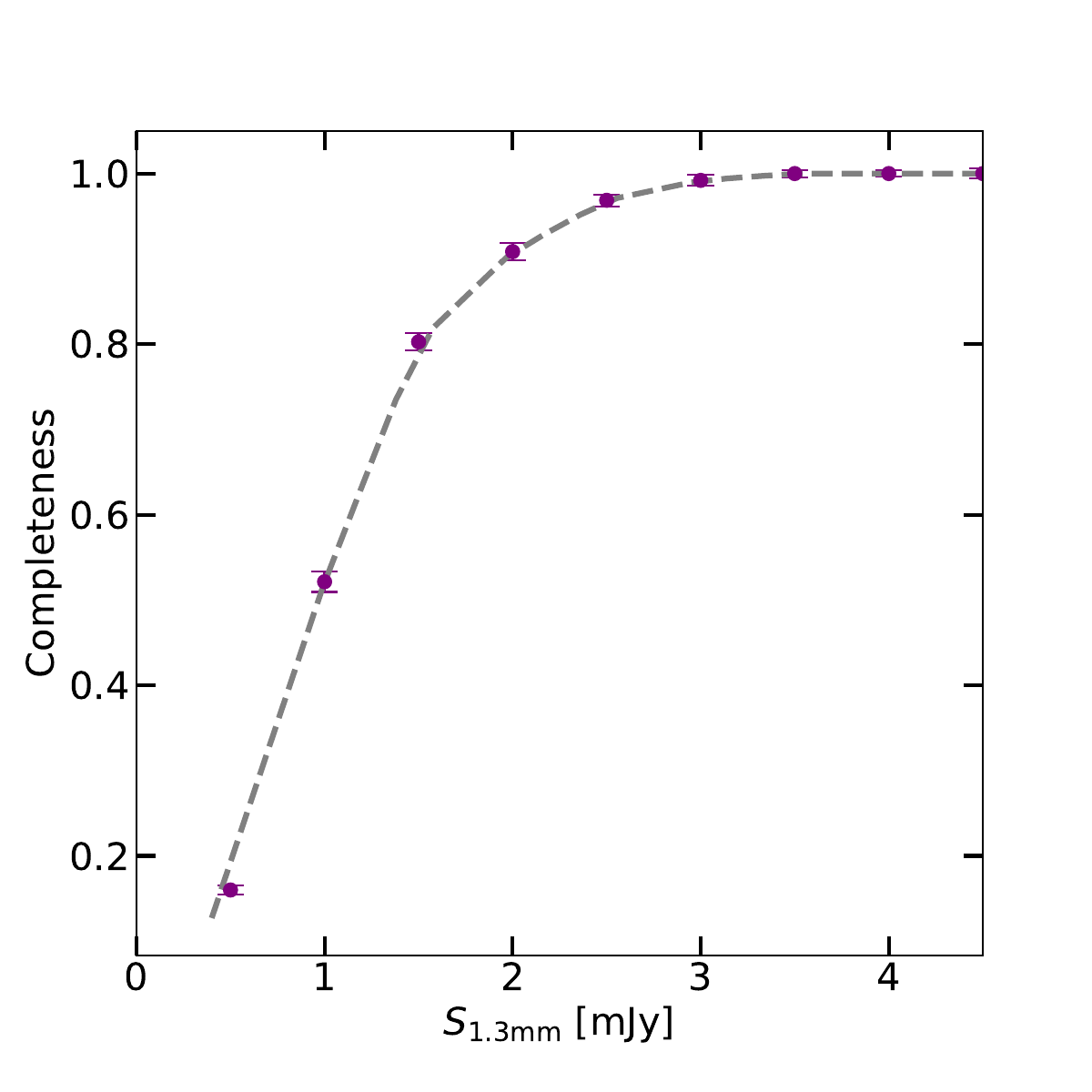}
    \caption{Completeness of the HARPAS catalogue, based on the injection of $\sim$4,000 simulated sources with flux densities $S_{1.3\mathrm{mm}}>0.4$\,mJy, equivalent to the lower limit of the HARPAS sample, into the observed ALMA maps corrected for the primary beam. Completeness reaches 100\,per\,cent for flux densities above 3.5\,mJy. The grey line represents the interpolation between bins used to estimate completeness.}
    \label{fig:completeness}
\end{figure}

We estimate the completeness of our sample at 1.3 mm by randomly injecting approximately 4,000 simulated sources with flux densities $S_{1.3\mathrm{mm}}>0.4$\,mJy, equivalent to the lower limit of our HARPAS sample, into the observed ALMA maps corrected for the primary beam. At each source position, we estimate the r.m.s. over a 1\,arcsecond radius area and determine its detectability using a 5$\sigma$ threshold. To quantify uncertainties, we repeat this process 100 times and compute the standard deviation. Figure \ref{fig:completeness} shows the completeness of the HARPAS sample. We find that it reaches 100\,per\,cent for flux densities above 3.5\,mJy. However, for flux densities of $\sim$2.7\,mJy, i.e. the median flux density of the M-Fields, the HARPAS catalogue is $>$90\,per\,cent complete.

We incorporate the completeness of the HARPAS catalogue (Fig. \ref{fig:completeness}) to our simulated ALMA pointigs in Sec. \ref{sec:sim} by generating a random number between 0 and 1 for each source in a simulated pointing. If the number is less than the completeness at the respective source flux density, the source is considered a detection; otherwise, it is discarded.

\bsp	
\label{lastpage}
\end{document}